\documentstyle[aaspp4,epsfig,12pt]{article}
\def\be{\begin{equation}}
\def\ee{\end{equation}}
\def\baray{\begin{eqnarray}}
\def\earay{\end{eqnarray}}
\def\Vvec{{\mbox{\boldmath $V$}}}
\def\Lvec{{\mbox{\boldmath $L$}}}

\def\Lhat{{\mbox{\boldmath $\hat L$}}}

\def\bhat{{\mbox{\boldmath $\hat b$}}}

\def\Omvec{{\mbox{\boldmath $\Omega$}}}
\def\Omhat{{\mbox{\boldmath $\hat\Omega$}}}
\def\omo{\Omhat^{(0)}}
\def\Sighat{{\mbox{\boldmath $\hat\Sigma$}}}
\def\omvec{{\mbox{\boldmath $\omega$}}}

\def\lamvec{{\mbox{\boldmath $\hat\lambda$}}}
\def\lamvecmu{\lamvec^\mu}

\def\rvec{{\mbox{\boldmath $r$}}}
\def\rhat{{\mbox{\boldmath $\hat r$}}}
\def\dotprod{{\mbox{\boldmath $\cdot$}}}
\def\crossprod{{\mbox{\boldmath $\times$}}}
\def\ehat{{\mbox{\boldmath $\hat e$}}}
\def\ehatz{{\mbox{\boldmath $\hat e_z$}}}
\def\ehatplus{{\mbox{\boldmath $\hat e_+$}}}
\def\ehatminus{{\mbox{\boldmath $\hat e_-$}}}
\def\ehatpm{{\mbox{\boldmath $\hat e_\pm$}}}
\def\khat{{\mbox{\boldmath $\hat k$}}}
\def\ehatone{\ehat_{\bf 1}}
\def\ehattwo{\ehat_{\bf 2}}
\def\ehatthree{\ehat_{\bf 3}}
\def\etil{{\tilde e}}
\def\Ctil{{\tilde C}}

\def\Ftil{{\tilde F}}
\def\msun{M_\odot}
\def\Tr{{\rm Tr}}
\def\Ibar{{\overline{I}}}

\def\delI{\delta I}
\def\cn{{\rm cn}}
\def\sn{{\rm sn}}
\def\dn{{\rm dn}}
\def\sineta{{\Omhat_-}}
\def\coseta{{\Omhat_+}}
\def\cosigma{{\Sighat_+}}
\def\sinsigma{{\Sighat_-}}
\def\onefac{{1+2\Delta(2\Omhat_+^2-1)+\Delta^2}}
\def\twofac{{{1+{\sn^2(\tau)(\Delta-\Delta_0)(1+\Delta)\over(\Delta+\Delta_0)(1-\Delta)}}}}
\def\onehat{{\hat 1}}
\def\twohat{{\hat 2}}
\def\threehat{{\hat 3}}
\def\ktil{{\tilde k}}
\def\tautil{{\tilde\tau}}
\def\tosc{\Delta t_{osc}}
\begin{document}
\title{Precession of Isolated Neutron Stars II:\\
Magnetic Fields and Type II Superconductivity}
\author{Ira Wasserman\\Center for Radiophysics and Space Research\\Cornell University, Ithaca, NY 14853
%\\(DRAFT -- NOT FOR CIRCULATION)
}

%\title{Free Precession of Magnetized Neutron Stars:\\ Evidence for Type II Superconductivity?}
%\author{Ira Wasserman\\Center for Radiophysics and Space Research\\Cornell University, Ithaca, NY 14853
%\\(DRAFT -- NOT FOR CIRCULATION)}
\date{\today}

\begin{abstract}

We consider the physics of free precession of a rotating neutron star with
an oblique magnetic field. We show that if the magnetic stresses are large
enough, then there is no possibility of steady rotation, and precession is
inevitable. Even if the magnetic stresses are not strong enough to prevent
steady rotation, we show that the minimum energy state is one in which the
star precesses. Since the moment of inertia tensor is inherently triaxial
in a magnetic star, the precession is periodic but not sinusoidal in time,
in agreement with observations of PSR 1828-11. However, the problem we
consider is {\it not} just precession of a triaxial body. If magnetic
stresses dominate, the amplitude of the precession is not set just by the angle
between the rotational angular velocity and any principal axis, which allows
it to be small without suppressing oscillations of
timing residuals at harmonics of the precession frequency. We argue that magnetic
distortions can lead to oscillations of timing residuals of the amplitude,
period, and relative strength of harmonics observed in PSR 1828-11
if magnetic stresses in its core are about 200 times larger than the classical
Maxwell value for its dipole field, and the stellar distortion induced by
these enhanced magnetic stresses is about 100-1000 times larger than the deformation
of the neutron star's crust. Magnetic stresses this large can arise if the
core is a Type II superconductor, or from toroidal fields $\sim 10^{14}$ G
if the core is a normal conductor. The observations
of PSR 1828-11 appear to require that the neutron star is slightly
prolate.

\end{abstract}

\section{Introduction}
\label{intro}

The convincing observation of free precession of PSR 1828-11 
(Stairs, Lyne \& Shemar 2000) poses challenges for
theories of neutron stars. Shaham (1977; 1986) argued that vortex line pinning
in the neutron star crust should prevent long term precession. Sedrakian, Wasserman
\& Cordes (1999) showed that precession is still prevented if vortex lines are not
pinned perfectly but vortex drag is strong. They also showed that even if vortex
drag is weak, precession is damped away. Link \& Cutler (2002) estimated the strength of
vortex line pinning forces, and argued that PSR 1828-11
may precess at large enough amplitude to unpin superfluid vortices in the crust. If
vortex drag is small this would remove one impediment to free precession, although
the free precession would still damp away eventually if it is not excited continuously. 

Here, we consider an additional feature of radiopulsars like PSR 1828-11, namely,
that they are strongly magnetized, with magnetic axes that are at an angle to their
rotation axes. Based on earlier work on magnetic stars by Mestel and collaborators
(Mestel \& Takhar 1972, Mestel et al. 1981, Nittman \& Wood 1981; see also Spitzer 1958),
we argue that precession may be {\it required} -- there is no equilibrium
corresponding to solid body rotation without precession for a rotating star with
an oblique magnetic field. For a fluid star, though,
we shall see that although the fluid must precess, the magnetic axis rotates
uniformly. Although Mestel et al. (Mestel \& Takhar 1972, Mestel et al. 1981, 
Nittman \& Wood 1981)
showed that hydrostatic balance also requires
fluid motions in addition to the precession which affect the stellar magnetic field,
these are too slow to be important observationally. 

Once we also take account of the solid crust of a neutron star in addition to
its fluid interior, we show that if the stellar distortions due to the magnetic
field are larger than the distortion of the crust, then steady state rotation
is very unlikely (but not necessarily impossible). If the magnetic stresses
inside PSR 1828-11 are simply due to the classical Maxwell stress tensor,
evaluated with the inferred dipole magnetic field strength, 
then they are too weak to require precession at a period $\sim 1$ year.
However, if the interior of the neutron star consists of a Type II superconductor,
the effective stress tensor is larger
than for a classical magnetic field,
according to Jones (1975) and Easson \& Pethick (1977); as was
emphasized by these authors,
and Cutler (2002), the
magnetic distortion is correspondingly larger. For PSR 1828-11, we estimate
that the distortion that would result in a core that contains a Type II
superconductor can lead to precession at a period of order 1 year.
A sufficiently strong toroidal magnetic field ($B_t\sim 10^{14}$ G) could also
lead to precession at this period without Type II superconductivity (see
e.g. Eq. [2.4] in Cutler [2002] with $B_c\to 0$).

Even if magnetic stresses are not strong enough to prevent steady rotation,
magnetic distortion in an oblique rotator alters the physics of free precession
qualitatively and quantitatively compared to what one would expect for
precession due to axisymmetric crustal distortions. Even if the crust is
axisymmetric, misalignment between the symmetry axis of the crust and the
magnetic field make the effective stellar moment of inertia inherently
triaxial. From a phenomenological viewpoint, one manifestation of this
loss of axisymmetry is that although the angular velocity of the star
is a {\it periodic} function of time in the frame rotating with the crust and
magnetic field, it is not a {\it sinusoidal} function of time. This feature
is consistent with observations of PSR 1828-11, which reveal behavior at
several different harmonically related frequencies (Stairs, Lyne \& Shemar
2000). The relative strengths
of the harmonics depend on the degree of nonaxisymmetry, and, presuming
an axisymmetric crust, on the distortions induced by the magnetic field.
For the ordinary Maxwell stresses evaluated just with the inferred dipole
field strength, the nonaxisymmetry would be small, but
distortions resulting from a Type II superconductor, or from a normal core
with a large toroidal field, could produce sufficient
nonaxisymmetry to lead to comparable amplitudes for at least the first
few harmonics of the fundamental precession period, as is observed.

Periodic, but not sinusoidal, precession would also arise if the neutron
star crust were simply nonaxisymmetric even if magnetic stresses were
negligible. Thus, the detection of harmonic behavior in PSR 1828-11
cannot, by itself, be taken to be evidence for 
amplified magnetic stresses in its core.
%Type II superconductivity
%or strong (normal) toroidal fields.
However, standard results for triaxial precession, which are reproduced
as a byproduct of the calculations we present, show that the oscillations
at harmonics of the precession frequency are smaller in amplitude than
the oscillation at the fundamental frequency by powers
of the precession amplitude. We shall see that this is not the case
in models where magnetic stresses predominate: the amplitudes of
oscillations at the precessiom frequency and twice the precession
frequency may be comparable {\it even at small amplitude}. 

Moreover, it is well known that the minimum energy state for rotation of
a nonaxisymmetric body is one in which the angular velocity and
angular momentum are aligned with the principal axis of the body
with largest eigenvalue of the moment of inertia tensor. There is
no precession at all in this state. However, we shall see that even
when magnetic stresses are not strong enough to {\it require} free
precession, the minimum energy state for an oblique rotator does
{\it not} correspond to alignment of the angular velocity with any
principal axis of the effective moment of inertia tensor. Thus,
the minimum energy state is one in which the star precesses.

We shall see that the timing residuals associated with the precession
can account for observations of long term oscillations in PSR 1828-11,
but that the explanation only works if the core of the neutron star
has sufficiently strong magnetic stresses that magnetic distortions
are 100-1000 times larger than crustal deformations.
%dominate. This can happen for either a Type II superconducting
%core (Jones 1975, Easson \& Pethick 1977), or if there is a substantial
%(normal) toroidal field in the stellar interior ($B_t\sim 10^{14}$ G; Cutler 2002).
%Otherwise, the magnetic stresses are 
%too weak to account for the amplitude and period of the observed
%wobbles. 
Thus, we argue that observations of free precession
in PSR 1828-11 offer evidence that the pulsar is in the regime where
magnetic stresses dominate, because either its core is a Type II
superconductor, or, if it is a normal conductor,
possesses a very large toroidal field.

We review the arguments given by Mestel and collaborators
(Mestel \& Takhar 1972, Mestel et al. 1981, Nittman \& Wood 1981)
in \S \ref{mestel}. In \S \ref{inevitable}, we consider the conditions
under which precession of an oblique rotator is {\it inevitable}.
In \S \ref{precess}, we consider the modified precession problem
for an axisymmetric crust and misaligned magnetic field. There we
show that the minimum energy state at a given angular momentum
is one in which the star precesses, provided that the magnetic
field is not either along or perpendicular to the symmetry axis
of the crust. There, we also review classical results for free
precession of triaxial bodies; we shall find that there are
three distinct cases of interest for a star where magnetic
distortions are important. We present limiting results for
the timing residuals expected in this model in \S \ref{pulsarrivaltimes},
and obtain approximate results for the limit in which crustal distortions
dominate in \S \ref{pulsibsmall}, and for the opposite limit in
which magnetic distortions dominate in \S \ref{pulsiblarge}.

Here, our main purpose is to present arguments that an oblique
rotator must precess. In \S \ref{application}, though, we 
present brief application of the 
model to observations of PSR 1828-11. There we argue that only
models in which magnetic distortions dominate can account for
all of the observed features of the long-term periodic timing
residuals from this pulsar. We also suggest that the data
favor prolate rather than oblate magnetic distortions.

Several appendices present cumbersome mathematical details
needed to derive (and verify!) analytic results presented in
the text.

\section{Review of Structure of Fluid Equilibria with Oblique Magnetic Fields}
\label{mestel}

Let us begin by reviewing the theory of rotating stars with oblique magnetic
fields developed by Mestel and collaborators (Mestel \& Takhar 1972, 
Mestel et al. 1981, Nittman \& Wood 1981; see also
Spitzer 1958). Consider a (fluid) star with angular
momentun $\Lvec=L\Lhat$. In the absence of a magnetic field, the angular velocity
of the star would be $\Omvec_0=\Lvec/I\equiv\Omega_0\Lhat$.
Let us work in a reference system rotating at a rate $\omvec=\omega\Lhat$, where
$\omega$ is unknown, but must be $\Omega_0$ to lowest order in small quantities.
Assume that the magnetic field is axisymmetric about an axis $\bhat$ that is
fixed in this frame; we can verify that this works out correctly to lowest
order later.

Solve for the density field of the star under the assumption that it is
stationary; Mestel et al. (Mestel \& Takhar 1972, Mestel et al. 1981, 
Nittman \& Wood 1981)
discuss the sizes of time dependent correction terms,
which are smaller than any we retain here. The result is a superposition
of three components assuming slow rotation and weak magnetic fields:
\be
\rho(\rvec)=\rho_0(r)+\rho_\omega(r,\rhat\dotprod\Lhat)
+\rho_B(r,\rhat\dotprod\bhat)~.
\label{rhoexpand}
\ee
Here, $\rho_0(r)$ is the spherical profile of the undistorted star, and
$\rho_\omega(r,\rhat\dotprod\Lhat)$ and $\rho_B(r,\rhat\dotprod\bhat)$
are the distortions due to rotation and magnetic fields, axisymmetric
about $\Lhat$ and $\bhat$, respectively. In absolute magnitude, the
rotational distortion is of order $\omega^2R^3/GM$ for a star of radius
$R$ and mass $M$, and the magnetic distortion is of order $BHR^4/GM^2$,  
where $H=B$ for ordinary magnetic fields, but $H$ corresponds to the first
critical field strength in a Type II superconductor ($H_{c1}\sim 10^{15}$G), 
as discussed by Jones (1975), Easson \& Pethick (1977) and Cutler (2002).

The moment of
inertia tensor corresponding to Eq. (\ref{rhoexpand}) is of the form
\be
I_{ij}=\int{d^3r~\rho(\rvec)\left(r^2\delta_{ij}-r_ir_j\right)}
=I_0\delta_{ij}+I_\omega\left(\delta_{ij}-3\Lhat_i\Lhat_j\right)
+I_B\left(\delta_{ij}-3\bhat_i\bhat_j\right)~,
\label{itensor}
\ee
where ($P_2(\mu)={1\over 2}\left(3\mu^2-1\right)$)
\baray
I_0&=&{2\over 3}\int{d^3r~\rho(\rvec)r^2}\nonumber\\
I_\omega&=&{1\over 3}\int{d^3r~\rho_\omega(r,\rhat\dotprod\Lhat)
P_2(\rhat\dotprod\Lhat)}\nonumber\\
I_B&=&{1\over 3}\int{d^3r~\rho_B(r,\rhat\dotprod\bhat) 
P_2(\rhat\dotprod\bhat)~.}
\label{ifacts}
\earay
Let $\Omvec=\Lvec/I+\delta\Omvec\equiv
\Omvec_0+\delta\Omvec$;
then the angular momentum of the star is
\be
\Lvec\equiv I\Omvec_0=(I_0+I_\omega+I_B)\Omvec-3I_\omega\Lhat\Lhat\dotprod\Omvec
-3I_B\bhat\bhat\dotprod\Omvec\simeq
(I_0-2I_\omega+I_B)\Omvec_0+I\delta\Omvec-3I_B\bhat\bhat\dotprod\Omvec_0~,
\ee
which implies that
\be
\delta\Omvec\simeq \left(1-{I_0\over I}+{2I_\omega\over I}-{I_B\over I}\right)
\Omvec_0+{3I_B\bhat\bhat\dotprod\Omvec_0\over I}~.
\ee
We can decompose this into components along and perpedicular to $\Lhat$ i.e.
$\delta\Omvec=\delta\Omega_\parallel\Lhat+\delta\Omvec_\perp$, where
\baray
\delta\Omega_\parallel&\simeq&\Omega_0\left[1-{I_0\over I}+{2I_\omega\over I}
+{2I_B\over I}P_2(\bhat\dotprod\Lhat)\right]\nonumber\\
\delta\Omvec_\perp&\simeq&{3I_B\Omega_0(\bhat\dotprod\Lhat)\over I}
\left(\bhat-{\Lhat\bhat\dotprod\Lhat}\right)
={3I_B\Omega_0(\bhat\dotprod\Lhat)[\Lhat\crossprod(\bhat\crossprod\Lhat)]\over
I}~;
\earay
$\delta\Omvec_\perp=0$ only if $\bhat$ is either parallel to or perpendicular
to $\Lhat$. Since Euler's equations imply time independent $\Omvec$ only if
$\Lvec$ and $\Omvec$ are parallel, in general a rotating star with an oblique
magnetic field must precess.

To find the angular velocity of the reference system, require that the
magnetic axis rotates with angular velocity 
$\Omvec\equiv\omega\Lhat+3\bhat I_B\bhat\dotprod\Omvec_0/I$; thus
\be
\omega=\Omega_0\left(2-{I_0\over I}+{2I_\omega\over I}-{I_B\over I}\right)
\ee
In this reference system, matter precesses about $\bhat$ with angular
velocity
\be
\omvec_p={3I_B\Omega_0\bhat\bhat\dotprod\Lhat\over I}~.
\label{precssfreq}
\ee
However, even though the fluid precesses, the magnetic field does not, so
we do not expect any observable effects to arise from the precession.

\section{Conditions Under Which Steady Rotation is Impossible and Precession
is Inevitable}
\label{inevitable}

We can look at the Mestel et al. (Mestel \& Takhar 1972, Mestel et al. 1981, 
Nittman \& Wood 1981)
results in a slightly different way, which
is identical to the same order of approximation, but differs in approach
slightly. Work in a reference system corotating with the matter.
\footnote{Mestel et al. (Mestel \& Takhar 1972, Mestel et al. 1981, Nittman \& Wood 1981)
show that maintaining hydrostatic balance also
requires additional motions of the fluid. We shall discuss this feature
in more detail below.}
The perfect conductivity condition demands that the magnetic axis is fixed in
this reference system. Instead of Eq.~(\ref{rhoexpand}), suppose that
\be
\rho(\rvec)=\rho_0(r)+\rho_\Omega(r,\rhat\dotprod\Omhat)
+\rho_B(r,\rhat\dotprod\bhat)~;  
\label{rhoexpand2}
\ee
in this case, the moment of inertia tensor (formerly Eq. [\ref{itensor}])
becomes
\be
I_{ij}=\int{d^3r~\rho(\rvec)\left(r^2\delta_{ij}-r_ir_j\right)}
=I_0\delta_{ij}+I_\Omega\left(\delta_{ij}-3\Omhat_i\Omhat_j\right)
+I_B\left(\delta_{ij}-3\bhat_i\bhat_j\right)~,
\label{itensor2}
\ee
where the various factors are defined just as in Eq. (\ref{ifacts}).
The angular momentum is therefore
\be
\Lvec=\left(I_0-2I_\Omega+I_B\right)\Omvec-3I_B\bhat\bhat\dotprod\Omvec.
\label{lvec2}
\ee
The condition for a time independent $\Omvec$ is $\Lvec\parallel\Omvec$.
This condition is satisfied only if $\bhat$ is either parallel to or
perpendicular to $\Omvec\simeq\Lhat$. The star precesses at an angular
frequency $\omvec_op$ as before. However, the magnetic axis rotates at
a uniform angular velocity, because, from Eq. (\ref{lvec2})
\be
\Omvec\crossprod\bhat={\Lvec\crossprod\bhat\over I_0-2I_\Omega+I_B}
\equiv\Omvec_B\crossprod\bhat~,
\ee
where
\be
\Omvec_B={\Lvec\over I_0+2I_\Omega+I_B}\simeq
\Omvec_0\left(2-{I_0\over I}+{2I_\Omega\over I}-{I_B\over I}\right)
\simeq\omvec~,
\ee
which is independent of time. 

The second approach makes contact with the Euler problem clearer. 
Angular momentum conservation is consistent with time independent
rotation as long as the angular velocity vector is aligned with
one of the principal axes of the moment of inertia tensor. For 
anisotropic density distribution, the moment of inertia tensor is
of the form
$$I_{ij}=I_0\delta_{ij}+\Delta I_{ij}$$
where $\Tr(\Delta I_{ij})=0$. Even though $\vert\vert\delta I_{ij}\vert
\vert\ll I_0$ for small distortions, the principal axes
of $I_{ij}$ are the principal axes of $\Delta I_{ij}$. For the magnetic
fluid, Eqs. (\ref{rhoexpand2}) and (\ref{itensor2}) imply that
$$\Delta I_{ij}=I_\Omega\left(\delta_{ij}-3\Omhat_i\Omhat_j\right)
+I_B\left(\delta_{ij}-3\bhat_i\bhat_j\right)~.$$
For there to be a principal axis of $\Delta_{ij}$ along $\Omvec$,
$$\Delta_{ij}\Omvec_j=\Lambda\Omvec_i~.$$
where $\Lambda$ is the associated eigenvalue. For this form of
$\Delta_{ij}$, the condition becomes
$$-3I_B\bhat_i\bhat\cdot\Omvec=(\Lambda+2I_\Omega-I_B)\Omvec_i,$$
which is only true in general provided that either $\Omvec$ is along
$\bhat$ or $\Omvec$ is perpendicular to $\bhat$, neither of which
will be the case generally.

Next, consider what happens when we consider a neutron star model
that consists of a rigid solid and a fluid, with an oblique
magnetic field. In that case, Eq. (\ref{itensor2}) is generalized
to \be
I_{ij}=I_0\delta_{ij}+C_{ij}+I_\omega(\delta_{ij}-3\hat\Omvec_i\hat
\Omvec_j)+I_B\left(\delta_{ij}-3\bhat_i\bhat_j\right)~,
\label{itensor3}
\ee
where $C_{ij}$ is the trace-free part of the moment of inertia of the crust.
\footnote{More precisely, $C_{ij}$ is the portion of the moment of inertia
tensor of the crust that is not aligned with the magnetic distortions. Any
crustal distortions symmetric about $\bhat$ would just renormalize 
$I_B$.}
The angular momentum of the star is therefore
\be
\Lvec_i=\left(I_0-2I_\Omega\right)\Omvec_i
+\left[C_{ij}+I_B\left(\delta_{ij}-3\bhat_i\bhat_j\right)\right]\Omega_j~.
\ee
A steady state is only possible if the star rotates along a principal axis of
$$\Delta I_{ij}\equiv C_{ij}+I_B\left(\delta_{ij}-3\bhat_i\bhat_j\right)~.$$
But for this to be true, we must have
\be
C_{ij}\Omvec_j-3\bhat_iI_B(\bhat\dotprod\Omvec)=(\Lambda-I_B)\Omvec_i~,
\ee
where $\Lambda$ is the associated eigenvalue. The components of this
equation perpendicular to $\Omvec$ are
\be
C_{ij}\Omhat_j-\Omhat_i\Omhat_kC_{kj}\Omhat_j
+3I_B(\bhat\dotprod\Omhat)\left[\Omhat_i(\bhat\dotprod\Omhat)-\bhat_i\right]
=0~.
\label{noprecess}
\ee
Eq. (\ref{noprecess}) must have a solution (with $\Omhat=\Lhat$) in
order for there to be no precession.

To understand the significance of Eq. (\ref{noprecess}), first review the
argument against precession when magnetic fields are ignored. The moment of
inertia tensor of the solid is not known observationally for any neutron star,
so we can specify $C_{ij}$ freely. Since it is a trace-free symmetric tensor,
$C_{ij}$ has five independent components. We only need to choose two of these
(the orientations of one of the three orthogonal principal axes) to get a
steady state. We can make these choices without considering the magnitude
of the distortion of the solid, since only the directions of the eigenvectors
matters for finding a steady state.
Moreover, from a physical viewpoint, we expect that at a fixed
$\Lvec$ the energy associated with rotation is minimized if $\Lvec$ (and
hence $\Omvec$) is along the principal axis with the largest eigenvalue;
dissipative processes tend to drive the neutron star toward
this state. 

Next, let us consider what happens when we restore the magnetic field.
Expand 
\be
C_{ij}=\sum_\mu C_\mu\lamvecmu_i\lamvecmu_j~,
\ee
where $C_\mu$ are the eigenvalues and $\lamvecmu$ the eigenvectors of
$C_{ij}$, and rewrite Eq. (\ref{noprecess}) as
\be
\sum_\mu C_\mu\left[\lamvecmu\dotprod\Omhat\lamvecmu-\Omhat
(\lamvecmu\dotprod\Omhat)^2\right]
+3I_B(\bhat\dotprod\Omhat)\left[\Omhat(\bhat\dotprod\Omhat)-\bhat\right]
=0~.
\label{noprecess2}
\ee
We know that if $I_B=0$, a steady state is possible irrespective of the
magnitudes $C_\mu$, and that if all $C_\mu=0$ there is no steady state.
When $I_B$ is nonzero, we can adjust the $\lamvecmu$ so that Eq.
(\ref{noprecess2}) is satisfied, just as for $I_B=0$, but the
adjustment depends on the magnitudes $C_\mu$ relative to $I_B$. 
In general, we would expect there to be no solution if $\vert C_\mu
\vert\lesssim I_B$, in which case the neutron star must precess.
In \S~\ref{precess} we shall also see that even when $I_B$ is not
large compared to $C$, the rotational energy is minimized, to first
order in distortions, when the angular velocity is not precisely
along a principal axis of the effective moment of inertia tensor,
and therefore not precisely along $\Lhat$.

To appreciate this result better quantitatively, let us focus on
the simplest special case in which the crust is axisymmetric. If
$\khat$ is the symmetry axis of the crust, then we may write
\be
C_{ij}=C\left(3\khat_i\khat_j-\delta_{ij}\right)
\ee
so that the largest eigenvalue of the moment of inertia of the crust
is along $\khat$. Then the condition for a steady state, 
Eq. (\ref{noprecess2}), may be written as
\be
3C(\khat\dotprod\Omhat)\left[\khat-\Omhat(\khat\dotprod\Omhat)\right]
+3I_B(\bhat\dotprod\Omhat)\left[\Omhat(\bhat\dotprod\Omhat)-\bhat\right]
=0~.
\label{noprecessaxi}
\ee
If Eq. (\ref{noprecessaxi}) has a solution, it must have coplanar
$\khat,\bhat\Omhat$. Thus, let us adopt $\bhat=\ehatthree$ and
\baray
\Omhat&=&\ehatthree\cos\chi+\ehatone\sin\chi\nonumber\\
\khat&=&\ehatthree\cos\theta+\ehatone\sin\theta~:
\label{angdef}
\earay
substitute into Eq. (\ref{noprecessaxi}) to find 
\be
\left(\ehatthree\sin\chi-\ehatone\cos\chi\right)
\left[C\cos(\chi-\theta)\sin(\chi-\theta)
-I_B\sin\chi\cos\chi\right]=0~.
\ee
Therefore, for a steady state, we must require
\be
\sin\left[2(\chi-\theta)\right]={I_B\sin 2\chi\over C}~,
\label{axianglesteady}
\ee
which is only possible if $\vert I_B\sin 2\chi\vert
\leq C$.  If $I_B\gtrsim C$, then steady state rotation is
rather unlikely, with a probability that decreases with
increasing $I_B/C$. When $I_B\sin 2\chi/C\leq 1$, Eq. (\ref{axianglesteady})
has the solutions
\baray
\cos 2\theta&=&{I_B\sin^22\chi\over C}\pm\cos 2\chi\sqrt{1-{I_B^2\sin^22\chi\over
C^2}}\nonumber\\
\sin 2\theta&=&\sin 2\chi\left(\pm\sqrt{1-{I_B^2\sin^22\chi\over C^2}}
-{I_B\cos 2\chi\over C}\right)~,
\earay
as can be verified by direct substitution; the choice of signs depends on
specific parameter values.

In order of magnitude, $3I_B/I\equiv\beta BHR^4/GM^2$, where $\beta\sim 1$ is a
structure constant, so that
\be
\omega_B\equiv {3I_B\Omega_0\cos\chi\over I}=1.9\times 10^{-9}\Omega_0
\beta\cos\chi (BH)_{27}R_6^4M_{1.4}^{-2}~,
\ee
for a neutron star mass and radius $M=1.4M_{1.4}\msun$ and $R=10R_6~{\rm km}$,
a magnetic axis inclination angle $\chi=\cos^{-1}(\bhat\dotprod\Lhat)$,
and $BH=10^{27}(BH)_{27}~{\rm G}$. ($BH\sim 10^{27}~{\rm G}$ for $B\sim 10^{12}
~{\rm G}$ and $H\simeq 10^{15}~{\rm G}\sim H_{c1}$ -- see Easson \& Pethick 1977.)
The period associated with $\omega_B$ is $P_B=2\pi/\omega_B
\simeq 16.6P_0({\rm s})(\beta\cos\chi)^{-1}
M_{1.4}^2R_6^{-4}(BH)_{27}^{-1}~{\rm y}$ for a neutron star rotation period $P_0
=1P_0({\rm s})~{\rm s}$. If magnetic effects are strong enough to
{\it require} precession, then we should
expect a precession period of order $P_B$. 
%Moreover, the conditions under which
%precession is required can only hold if $P_B$ is shorter to or comparable to
%the precession period. 
We consider the precession period more completely in
the next section.

For PSR 1828-11, the spin period is $P_0=0.405~{\rm s}$
and the dipole field strength deduced from the observed spindown 
is $B\simeq 5\times 10^{12}~{\rm G}$ (Stairs, Lyne \& Shemar 2000); these values
would imply $P_B\simeq 1.35(\beta\cos\chi)^{-1}[5/(BH)_{27}]~{\rm y}$, or about
$492(\beta\cos\chi)^{-1}(BH_{27}/5)~{\rm d}$, similar to the observed period
for $\beta\cos\chi\sim 1$. Note that the precession period would be far
longer for $BH\to B^2$, by a factor of about 200. Thus, if magnetic effects
are the reason for the observed ``precession'' then ether the neutron star interior
must be a Type II superconductor (Jones 1975, Easson \& Pethick 1977, Cutler 2002), 
or there must be a substantial toroidal field in the core if it is not a Type II
superconductor (e.g. Eq. [2.4] in Cutler 2002 with $B_c\to 0$).

\section{Modified Euler Problem}
\label{precess}

\subsection{Basic Equations, Principal Axes and Eigenvalues}
\label{eigen}

Next, let us consider the modified Euler problem. To allow an analytic
treatment, continue to assume that the crust is axisymmetric. Although this
is a simplification, the effective moment of inertia of the star will still
be triaxial because of the misaligned distortion introduced by the magnetic
field, except for special orientations.

The Euler equation in the rotating frame of reference is
\be
I^{eff}_{ij}{d^\star\Omega_j\over dt}+3\epsilon_{ijl}\Omega_j
\left(C\khat_l\khat_m-I_B\bhat_l\bhat_m\right)\Omega_m=0~,
\label{euler}
\ee
where the effective moment of inertia tensor is
\be
I^{eff}_{ij}=(I_0-2I_\Omega+I_B-C)\delta_{ij}
+3\left(C\khat_i\khat_j-I_B\bhat_i\bhat_j\right)~.
\label{ieffdef}
\ee
and $d^\star/dt$ is the time derivative in the rotating frame.
The two vectors, $\bhat$ and $\khat$ define a plane, and one of the
eigenvectors of $I^{eff}_{ij}$ is along the unit vector perpendicular
to that plane, $\ehattwo$, and has an eigenvalue $I_2=I_0-2I_\Omega+I_B-C$.
The other two eigenvectors lie in the $\bhat-\khat$ plane. As in the previos
section, take $\ehatthree=\bhat$ and define $\khat$ by Eq. (\ref{angdef}).
Then the other two eigenvectors, $\ehatpm$ are
\footnote{The following results imply that
$$\ehatplus\crossprod\ehatminus=\ehattwo$$
$$\ehattwo\crossprod\ehatplus=\ehatminus$$
$$-\ehattwo\crossprod\ehatminus=\ehatplus$$
so that these axes define a right handed coordinate system.}
\be
\ehatpm=\ehatthree\cos\sigma_\pm+\ehatone\sin\sigma_\pm~,
\ee
where 
\baray
\sin\sigma_\pm&=&{1\over\sqrt{2}}\left(1\pm{\eta\over\sqrt{1+\eta^2}}\right)^{1/2}
\nonumber\\
\cos\sigma_\pm&=&\pm{1\over\sqrt{2}}\left(1\mp{\eta\over\sqrt{1+\eta^2}}\right)^{1/2}
\nonumber\\
\eta&=&{I_B-C\cos 2\theta\over C\sin 2\theta}~.
\label{eigenvectors}
\earay
The eigenvalues associated with these eigenvectors are $I_\pm=I_2+\delta I_\pm$, where
\be
\delta I_\pm={3\over 2}\left[C-I_B\pm\sqrt{C^2-2CI_B\cos 2\theta+I_B^2}\right]~.
\label{eigenvalues}
\ee
In Appendix \ref{app:appendixa} we show that for $I_B>0$, $I_+>I_2>I_-$, and
for $I_B<0$, $I_+>I_->I_2$.

Approximate results are derived for $\vert I_B\vert\ll C$ in Appendix \ref{app:ibllc}
and for $\vert I_B\vert\gg C$ in Appendix \ref{app:ibggc}. When $\vert I_B\vert\ll C$,
the eigenvectors are nearly aligned with the principal axes of $C_{ij}$, and the
eigenvalues are nearly the eigenvalues of $C_{ij}$, apart from small corrections
$\sim\vert I_B/C\vert$. For $\vert I_B\vert\gg C$, the eigenvectors are nearly along and
perpendicular to $\bhat$ and the eigenvalues are almost determined by the magnetic
distortions alone, apart from corrections $\sim\vert C/I_B\vert$. We shall argue below
that the case $I_B<0$, $\vert I_B/C\vert\gg 1$ may be especially relevant to observations
of PSR 1828-11.

\subsection{Solution of the Euler Equations}
\label{solution}

\subsubsection{Basic Equations and Minimum Energy State}
\label{basics}

The modified Euler equations are simply what one finds in general for 
a triaxial system,
\baray
I_+{d^\star\Omega_+\over dt}-\delta I_-\Omega_2\Omega_-&=&0\nonumber\\
I_-{d^\star\Omega_-\over dt}+\delta I_+\Omega_2\Omega_+&=&0\nonumber\\
I_2{d^\star\Omega_2\over dt}-(\delta I_+-\delta I_-)\Omega_+\Omega_-
&=&0~.
\label{eulercomps}
\earay
As is well-known, Eqs. (\ref{eulercomps}) conserve both the magnitude of $\Lvec$ and
the rotational energy,
\be
E_{rot}={1\over 2}\left(I_+\Omega_+^2+I_-\Omega_-^2+I_2\Omega_2^2\right)~.
\ee
Note that the magnitude of the angular velocity is not conserved, so that
in actuality $I_\Omega$ is not independent of time. However we will assume
that the variation is slow enough, and of small enough amplitude, that its
effect is only higher order than any others we consider here.

The condition for steady rotation, Eq. (\ref{axianglesteady}),
can be rederived from Eqs. (\ref{eigenvectors}) under the assumption that
$\Omhat$ is along either $\ehatplus$ or $\ehatminus$, with $\chi<\pi/2$ or
$\chi>\pi/2$, respectively. We are interested in what happens when precession
is required, so let us assume that Eq. (\ref{axianglesteady}) cannot be satisfied
for any choice of $\theta$ given $\chi$, $I_B$ and $C$. Nevertheless, we expect
that the system seeks a minimum energy state, even if that state is not steady.
To be specific, let $\chi$ be the angle between $\bhat$ and $\Omhat$ when
they are coplanar with $\khat$ (and therefore with $\ehatpm$). Let us
determine $\theta$ from the requirement that $E$ is minimum for a given
$L^2$. Thus, let us consider the quantity $2E/L^2$ at the epoch when
$\Omhat$, $\khat$ and $\bhat$ are coplanar. If we define
\baray
\Ibar&\equiv&{1\over 2}(I_++I_-)=I_2+{3(C-I_B)\over 2}\nonumber\\
\delI&\equiv&{1\over 2}(I_+-I_-)\nonumber\\
\Delta&\equiv&{\delI\over\Ibar}={3\sqrt{C^2-2CI_B\cos 2\theta+I_B^2}\over
2I_2+3(C-I_B)}~,
\label{ibardeltadef}
\earay
then we find
\be
{2\Ibar E\over L^2}={1+\Delta(2\Omhat_+^2-1)\over 1+\Delta^2+2\Delta
(2\Omhat_+^2-1)}\simeq 1-\Delta(2\Omhat_+^2-1)~,
\label{energy}
\ee
where the approximation holds for $\Delta\ll 1$, and
\baray
\Omhat_+\equiv\Omhat\dotprod\ehatplus
&=&{1\over\sqrt{2}}\biggl\{\cos\chi\left[1-{(I_B-C\cos 2\theta)
\over\sqrt{C^2-2CI_B\cos 2\theta+I_B^2}}\right]^{1/2}
\nonumber\\& &
+\sin\chi\left[1+{(I_B-C\cos 2\theta)
\over\sqrt{C^2-2CI_B\cos 2\theta+I_B^2}}\right]^{1/2}\biggr\}\nonumber\\
\Omhat_-\equiv\Omhat\dotprod\ehatminus
&=&{1\over\sqrt{2}}\biggl\{\sin\chi\left[1-{(I_B-C\cos 2\theta)
\over\sqrt{I_B^2-2CI_B\cos 2\theta+C^2}}\right]^{1/2}
\nonumber\\& &
-\cos\chi\left[1+{(I_B-C\cos 2\theta)\over
\sqrt{I_B^2-2CI_B\cos 2\theta+C^2}}\right]^{1/2}\biggr\}~.
\label{omhatpm}
\earay
Assuming that $\theta\leq\pi/2$ we find
\be
\Delta(2\Omhat_+^2-1)={3\left\{C\cos[2(\chi-\theta)]-I_B\cos 2\chi\right\}\over
2I_2+3(C-I_B)}~.
\ee
To first order in $\Delta$, the energy is minimized when
$\Delta(2\Omhat^2-1)$ is maximized, which happens when $\theta=\chi$.

Note that in the minimum energy configuration,
$\Omhat_-$ need not be very small, although it vanishes
in the minimum energy state for $I_B\to 0$. For small values of
$I_B/C$, 
\be
\Omhat_-\simeq \sin(\chi-\theta)-{I_B\over 2C}\sin 2\theta
\cos(\chi-\theta)\to -{I_B\sin 2\theta\over 2C}~,
\ee
where the last result assumes $\theta\to\chi$. Thus, for small
$I_B$, we find small but nonzero $\Omhat_-$ in the minimum
energy state. For $\vert I_B/C\cos 2\theta\vert\gg 1$,
$\Omhat_-\simeq -I_B\sin 2\chi/2C$. On the other hand,
for $\vert I_B/C\cos 2\theta\vert
\gg 1$, $\Omhat_-\simeq
\sin\chi$ or $-\cos\chi$, depending on the sign of $I_B$.

When $I_B\equiv 0$, $\theta=\chi$ minimizes $2E\Ibar/L^2$ exactly, and
we find, as usual, $\Omhat_+=1$, corresponding to an angular velocity
aligned with the principal axis with the largest moment of inertia.
When $I_B\neq 0$, $\theta=\chi$ does not correspond to exact alignment
of the angular velocity and the principal axis with the largest moment
of inertia. This is because the eigenvalues of $I^{eff}_{ij}$
depend on $\theta$, so $\Delta$ depends on $\theta$. If $\Delta$
were independent of $\theta$, then the energy would be minimized for $\Omhat_+=1$,
the largest possible value of $\Omhat_+$. The value of 
\be
\Lhat\dotprod\Omhat={1+\Delta(2\Omhat_+^2-1)\over
\sqrt{1+2\Delta(2\Omhat_+^2-1)+\Delta^2}}\simeq 1-2\Delta^2\Omhat_+^2\Omhat_-^2
\ee
is only exactly one when either $\Omhat_+=1$ or $\Omhat_-=1$; the angle between
$\Lhat$ and $\Omhat$ is $\pm 2\Delta\Omhat_+\Omhat_-$ more generally.

The condition that $\theta=\chi$ means that the angle between the angular
velocity vector and the magnetic field is the same as the angle between
the symmetry axis of the crust and the field. Thus, it corresponds to
angular velocity that is parallel to the symmetry axis of the crust
at the epoch when all of the vectors lie in a plane. Since the angle
between the angular velocity vector and the magnetic axis is fixed in
this case, the portion of the rotational energy due to the rotating
magnetic distortion is fixed. Minimizing the energy then amounts to 
minimizing the portion of the energy associated with rotation of the
crustal distortions. This is achieved if the angular velocity is along
the symmetry axis of the crust. The resulting minimum is only a local
minimum, at a given value of $\chi$. The global minimum is achieved
for $\chi=0$ or $\pi/2$. We assume that the star can evolve on a slow
disspative timescale toward the local minimum, and on a longer timescale,
probably the spindown timescale (e.g. Goldreich 1970) toward the
global minimum.

The fact that $\Delta=\Delta(\cos 2\theta)$
when $I_B\neq 0$ implies that even when alignment of $\Omvec$ and 
$\Lvec$ is possible, the minimum energy state is not steady rotation,
but rather precession. When $I_B\ll C$, the minimum energy state corresponds
to an angle of approximately $\vert\Omhat_-\vert
\simeq\vert I_B\sin 2\chi\vert/2C$ between $\Omhat$ and $\ehatplus$,
and an angle $\simeq 3\vert I_B\sin 2\chi\vert/2\Ibar$ between
$\Omhat$ and $\Lhat$.

\subsubsection{Nonlinear Solution: $I_B>0$}
\label{nlibpos}

The complete, nonlinear solution to Eqs. (\ref{eulercomps}) is given in
Landau \& Lifshitz, \S37, in terms of elliptic functions. We will have
to consider separately the two cases $I_B>0$ and $I_B<0$. Here, we 
consider $I_B>0$, which implies $I_+>I_2>I_-$ according to
Eq. (\ref{inequalpos}) and the ensuing discussion. Adapting the
solution in Landau \& Lifshitz, \S37, to this situation (and our
notation) we have (see their Eqs. [37.8]-[37.12])
\footnote{There is actually a sign ambiguity in the solution
given by Landau \& Lifshitz, which we resolve by always choosing
$\Omega_-\propto\Omhat_-$, rather than $\Omega_-\propto\sqrt
{1-\Omhat_+^2}$. The Euler equations Eq. (\ref{euler}) require
that if we choose $\Omega_-\propto\Omhat_-$, then we should also
choose the sign of the coefficient of $\Omega_2$ to be
the same as the sign of the coefficient of $\Omega_-$.}
\baray
{\Ibar\Omega_-\over L}&=&{\Omhat_-\cn(\tau)\over
\sqrt{\onefac}}
%\simeq\cn(\tau)\sqrt{1-\Omhat_+^2}
\nonumber\\
{\Ibar\Omega_2\over L}&=&\Omhat_-\sn(\tau)\sqrt{2\Delta(1-\Delta)\over
[1+2\Delta(2\Omhat_+^2-1)+\Delta^2]
(1-\Delta_0)(\Delta+\Delta_0)}
%\simeq\sn(\tau)\sqrt{2\Delta(1-\Omhat_+^2)\over
%\Delta+\Delta_0}
\nonumber\\
{\Ibar\Omega_+\over L}&=&{\Omhat_+~\dn(\tau)\over\sqrt{1+2\Delta(2\Omhat_+^2-1)
+\Delta^2}}\simeq\Omhat_+\dn(\tau)~,
\label{nlsolibpos}
\earay
where $\Delta_0\equiv 3(C-I_B)/2\Ibar$, $\cn(\tau)\equiv \sqrt{1-\sn^2(\tau)}$,
$\dn(\tau)\equiv \sqrt{1-k^2\sn^2(\tau)}$, $\sn(\tau)$ is defined by
\be
\tau=\int_0^{\sn(\tau)}{dx\over\sqrt{(1-x^2)(1-k^2x^2)}},
\label{sndef}
\ee
with
\baray
k^2&=&{(\Delta-\Delta_0)(1-\Delta)(1-\Omhat_+^2)
\over (\Delta+\Delta_0)(1+\Delta)\Omhat_+^2}
\simeq {(\Delta-\Delta_0)(1-\Omhat_+^2)\over(\Delta+\Delta_0)\Omhat_+^2}\nonumber\\
{d\tau\over dt}&=&{L\left\vert\Omhat_+\right\vert\over\Ibar}
\sqrt{2\Delta(1+\Delta)(\Delta+\Delta_0)
\over(1-\Delta_0)(1-\Delta^2)[1+2\Delta(2\Omhat_+^2-1)+\Delta^2]}
\nonumber\\
%&\simeq&{L\left\vert\Omhat_+\right\vert t\over\Ibar}\sqrt{2\Delta(\Delta+\Delta_0)}~.
\label{ktaudefpos}
\earay
%The approximate results in Eqs. (\ref{nlsolibpos}) and (\ref{ktaudefpos}) are correct to
%lowest order in $\Delta$ and $\Delta_0$.
The motion is periodic, with a dimensionless period $4K(k^2)$ where
\be
K(k^2)=\int_0^1{dx\over\sqrt{(1-x^2)(1-k^2x^2)}}~.
\ee
Eqs. (\ref{nlsolibpos}) and (\ref{ktaudefpos}) also imply that 
\be 
{\Ibar^2\vert\Omvec\vert^2\over L^2} 
={1-[2\Delta(\Delta-\Delta_0)(1-\Omhat_+^2)/(1+\Delta)(1-\Delta_0)]~\sn^2(\tau)\over 
1+2\Delta(2\Omhat_+^2-1)+\Delta^2}~, 
\label{omsqpos}
\ee 
which is independent of time up to terms $\sim\Delta^2$, thus validating the approximation 
of time independent $I_\Omega$ used above. 

One of the distinguishing features of the timing model we will develop below is that
we shall {\it not} demand that $\vert\Omhat_-\vert$ be small. In particular, we shall
see that when $\vert I_B\vert\gg C$, $\Omhat_-$ will not be small in general, but the
{\it observable} effect of the precession on pulse arrival times could still be small.
This is because in the limit where magnetic distortions are far larger than crustal
distortions the star tends to precess about its magnetic axis. If there were no
crust at all, as in the magnetic fluid stars considered by Mestel and collaborators
(Mestel \& Takhar 1972, Mestel et al. 1981, Nittman \& Wood 1981),
the star would precess exactly around its magnetic axis, which would therefore
rotate uniformly. The crust breaks this symmetry, and allows the precession to be
observable. We shall see this emerge in some detail when we consider timing residuals
in \S \ref{pulsiblarge}.

For the opposite case, where the crustal deformations dominate, the precession
amplitude is set by $\Omhat_-$, according to Eqs. (\ref{nlsolibpos}). In that case,
we also see that Eq. (\ref{ktaudefpos}) shows that $k^2\propto\Omhat_-^2$. Thus,
if the precession amplitude is small, so is $k^2$ irrespective of how triaxial the
crust might be. Since $k^2$ governs the importance of oscillations at harmonics
of the fundamental precession frequency, we see that small amplitude will imply
oscillations predominantly at the fundamental if crustal deformations dominate.
(This will also be true for the solution given in \S \ref{nlibneg} for 
$I_B<0$, where Eq. [\ref{ktaudefneg}] will also imply that oscillations at
harmonics of the precession frequency are suppressed for small precession
amplitude.)
This is not the case when magnetic deformations dominate.

The better-known results for free precession of an axisymmtric star are
recovered for $\Delta-\Delta_0=0=k^2$. However, note that in the triaxial
case, $k^2$ need not be small. We have already noted that $1-\Omhat_+^2$ 
does not have to be small, even in the minimum energy state, and from 
Eqs. (\ref{ibardeltadef}) and the definition of $\Delta_0$ we see that
\be
{\Delta-\Delta_0\over\Delta+\Delta_0}=
{\sqrt{C^2-2CI_B\cos 2\theta+I_B^2}-(C-I_B)\over
\sqrt{C^2-2CI_B\cos 2\theta+I_B^2}+C-I_B}~,
\ee
which is not necessarily small either. For $0<I_B/C\ll 1$, we note
that $k^2\simeq I_B^3\sin^22\chi\sin^2\chi/4C^3$ to lowest order in
$I_B/C$ in the minimum energy state, so $k^2\ll 1$ in this case.
However, for $I_B\gg C$, we shall
see that $k^2\sim I_B/C\gg 1$, and in general, since $\Delta_0<0$ for
$I_B>C$, it is possible for $k^2$ to exceed one rather generally in
that regime. The solution to the
Euler problem is still given in terms of elliptic equations when
$k^2>1$, but we have to make the replacements
\baray
\sn(\tau)&\to&{\sn(k\tau)\over k}\nonumber\\
\cn(\tau)&\to&\dn(k\tau)\nonumber\\
\dn(\tau)&\to&\cn(k\tau)
\label{kbigtransf}
\earay
in Eqs. (\ref{nlsolibpos}), and $\dn^2(k\tau)=1-k^{-2}\sn^2(k\tau)$. 
The explicit solution for $k^2>1$ is
\baray
{\Ibar\Omega_-\over L}&=&{\Omhat_-\dn(\tautil)\over\sqrt{\onefac}}
\nonumber\\
{\Ibar\Omega_2\over L}&=&\Omhat_+{\rm sign}(\Omhat_-)
\sn(\tautil)\sqrt{{2\Delta(1+\Delta)\over
(1-\Delta_0)(\Delta-\Delta_0)[\onefac]}}\nonumber\\
{\Ibar\Omega_+\over L}&=&{\Omhat_+\cn(\tautil)\over\sqrt{\onefac}}~,
\label{nlsolibposkbig}
\earay
where ${\rm sign}(\Omhat_-)$ is the sign of $\Omhat_-$,
which may be positive or negative,
$\dn^2(\tautil)\equiv 1-\ktil^2\sn^2(\tautil)$, and
\baray
{d\tautil\over dt}&=&{L\over\Ibar}\sqrt{{2\Delta(\Delta-\Delta_0)(1-\Omhat_+^2)
\over (1-\Delta_0)(1+\Delta)[\onefac]}}\nonumber\\
\ktil^2&=&{1\over k^2}={(\Delta+\Delta_0)(1+\Delta)\Omhat_+^2\over
(\Delta-\Delta_0)(1-\Delta)(1-\Omhat_+^2)}~.
\label{ktaudefibposkbig}
\earay
This somewhat unfamiliar case is not treated in Landau \& Lifshitz,
but is found eaasily from the equations given there, and corresponds
simply to the transformations in Eqs. (\ref{kbigtransf}).
Physically, it turns out to be important for a substantially
prolate figure, which is what happens when $I_B$ is large and
positive.

An important difference between axisymmetric and triaxial precession
is that even though both are periodic, only the axisymmetric case
is precisely sinusoidal. The functions $\cn(\tau)$ and $\sn(\tau)$
contain all {\it odd} harmonics of $\pi\tau/2K(k^2)$, whereas
the function $\dn(\tau)$ contains all {\it even} harmonics. When
$k^2$ is small, the expansions are dominated by their leading terms,
but, as we have seen, the general case does not demand small values
of $k^2$. The key parameter in the expansions is
\be
q(k^2)\equiv\exp\left[-{\pi K^\prime(k^2)\over K(k^2)}\right]~,
\ee
where
\be
K^\prime(k^2)=K(1-k^2)
\ee
(Abramowitz \& Stegun, Eqs. [16.1.1], [16.23.1-3]). Since $q(k^2)<1$ 
by its definition, only the first few harmonics ought to be prominent
in the solution, but as long as $k^2$ is not especially small 
compared to one, the amplitudes of the first few harmonics ought
to be roughly comparable. This feature is consistent with the 
observed properties of PSR 1828-11 (Stairs, Lyne \& Shemar
2000). If this interpretation of the
observations is correct, then the fundamental precession period
must be 1000 (or perhaps 2000) days. \footnote{The longer period appears
in the Fourier analysis of the timing residuals, but not in those
of the period, period derivative, or shape variations, given in
Fig. 2 of Stairs, Lyne \& Shemar 2000. In any event, the data
shown there only span about 2000 days, so this feature may not
be so well-established.}

\subsubsection{Nonlinear Solution: $I_B<0$}
\label{nlibneg}

For $I_B<0$, the solution is analogous to what was given in
\S \ref{nlibpos}, except that, according to Eq. (\ref{inequalneg})
and ensuing discussion, $I_+>I_->I_2$. Instead of Eqs. (\ref{nlsolibpos})
we have
\baray
{\Ibar\Omega_2\over L}&=&\Omhat_-\left\{{2\Delta(1-\Delta)\over
[\onefac](1-\Delta_0)(\Delta+\Delta_0)}\right\}^{1/2}\cn(\tau)
\nonumber\\
{\Ibar\Omega_-\over L}&=&\left[{\Omhat_-\sn(\tau)\over\sqrt{\onefac}}
\right]
\nonumber\\
{\Ibar\Omega_+\over L}&=&\left\{{[\Delta_0+\Delta^2+\Delta(1+\Delta_0)
(2\Omhat_+^2-1)]\over[\onefac](1+\Delta)(\Delta+\Delta_0)}\right\}^{1/2}
\dn(\tau)~,
\label{nlsolibneg}
\earay
and instead of Eq. (\ref{ktaudefpos}) we have
\baray
k^2&=&{(\Delta_0-\Delta)(1-\Delta)(1-\Omhat_+^2)\over
\Delta_0+\Delta^2+\Delta(1+\Delta_0)(2\Omhat_+^2-1)}
\nonumber\\
{d\tau\over dt}&=&{L\over\Ibar}\sqrt{{2\Delta[\Delta_0+\Delta^2
+\Delta(1+\Delta_0)(2\Omhat_+^2-1)]\over
(1-\Delta_0)(1-\Delta^2)[\onefac]}}~.
\label{ktaudefneg}
\earay
The definitions of $\sn(\tau)$, $\cn(\tau)$ and $\dn(\tau)$ are the
same as before. The magnitude of the angular velocity is now
\be
{\Ibar^2\vert\Omvec\vert^2\over L^2}=
{1+[2\Delta(\Delta_0-\Delta)(1-\Omhat_+^2)/(1+\Delta)(1-\Delta_0)]
\cn^2(\tau)\over\onefac}~.
\label{omsqneg}
\ee
Eq. (\ref{omsqneg}) implies that $\vert\Omvec\vert$ is independent
of time up to terms $\sim\Delta^2$, just as we found from Eq.
(\ref{omsqpos}) for $I_B>0$. In this case, $k^2<1$, so these solutions
suffice.

\subsection{Effect of the Spindown Torque}
\label{spindown}
  
The solutions given in \S \ref{nlibpos} and \ref{nlibneg} are for
{\it free} precession. Radiopulsars spin down as a result of electromagnetic
radiation, on a timescale $t_{sd}\equiv P/2\dot P$ that is long compared
with the precession periods of interest: for PSR 1828-11, $t_{sd}\simeq
1.1\times 10^5$ years. A careful examination of the Euler equations with
spindown included shows that the precession is described by the torque-free
solutions up to small corrections (just as was found by Link \& Epstein
2001). We also note here that the quantity $2E\Ibar/L^2$ only changes on
a still longer timescale, $\sim t_{sd}/\Delta$.
However, sinusoidal variations of the spindown torque result
from the precession, since the angle between $\bhat$ and $\Omvec$ varies
with time. The amplitude of these variations need not be small, and
the associated timing residuals can dominate (Cordes 1993). In fact, we shall
see that they are dominant, just as was found by Link \& Epstein (2001)
for precession of an axisymmetric neutron star.

For vacuum magnetic dipole radiation, 
\be
{d\Lvec\over dt}=K\bhat\crossprod(\bhat\crossprod\Omvec),
\ee
where $K$ is a constant to sufficient accuracy, and the magnitude of the
angular velocity changes at a rate
\be
{d\Omega\over dt}={1\over\Ibar}{dL\over dt}
\simeq -{\Omega_0\over 2t_{sd}}\left(1+{\zeta\over\sin^2\chi}\right)~,
\label{spindowntorque}
\ee
where $\Omega_0$ is the angular frequency at some reference time $t=0$;
in the absence of precession, $\zeta=0$. We use the conventional definition
of the spindown time as $t_{sd}=P/2\dot P=-\Omega/2\dot\Omega$.
We can use the solutions to the
Euler problem found above to find the time dependent $\zeta$ case by
case:
\baray
\zeta&=&\Omhat_-^2\sn^2(\tau)\left[{(\Delta-\Delta_0)\cos^2\sigma_+\over
\Delta+\Delta_0}+\sin^2\sigma_+\right]
-2\Omhat_+\Omhat_-\cos\sigma_+\sin\sigma_+[1-\dn(\tau)\cn(\tau)]
\nonumber\\
\zeta&=&\Omhat_+^2\sn^2(\tautil)\left[{(\Delta+\Delta_0)\sin^2\sigma_+\over
\Delta-\Delta_0}+\cos^2\sigma_+\right]
-2\Omhat_+\Omhat_-\sin\sigma_+\cos\sigma_+[1-\dn(\tautil)\cn(\tautil)]
\nonumber\\
\zeta&=&\Omhat_-^2\cn^2(\tau)\left[{(\Delta-\Delta_0)\cos^2\sigma_+\over
\Delta+\Delta_0}+\sin^2\sigma_+\right]
\nonumber\\& &
+2\Omhat_-\sin\sigma_+\cos\sigma_+\left[\dn(\tau)\sn(\tau)
\sqrt{\Delta_0+\Delta(2\Omhat_+^2-1)\over\Delta+\Delta_0}-
\Omhat_+\right]~,
\label{spindownvariation}
\earay
for, respectively, $I_B>0, k^2<1$, $I_B>0,\ktil^2=1/k^2<1$, and
$I_B<0$. Here, we do not restrict the solution to small values of
$\Omhat_-^2$, and Eqs. (\ref{spindowntorque}) and (\ref{spindownvariation})
imply oscillations of the spindown torque at both even and odd harmonics
of the precession frequency. (There are also evidently zero frequency
corrections but these can always be combined with $\sin^2\chi$ and factored out.)
Note that for simple triaxial precession
with a small tilt of the angular velocity away from the principal axis
with maximum moment of inertia, the amplitude of the oscillations at
$2\omega_p$ would be smaller, by a factor $\sim\vert\Omhat_-\vert$, than
the amplitude of the oscillations at $\omega_p$. For the model developed
here, $\Omhat_-$ can become substantial, particularly if $I_B<0$ and
$\vert I_B\vert\gg C$.

\section{Pulse Arrival Times}
\label{pulsarrivaltimes}

To determine pulse arrival times, we need to represent $\bhat(t)$
in the inertial frame of reference of the observer. We can define
the angular momentum vector $\Lvec$, which is conserved apart from
spindown in this reference frame, to lie along the $z$ axis, and
we can further choose to place the observer in the $x-z$ plane.
Pulses arrive when $\bhat_y(t)=0$ (actually, only half of the 
solutions correspond to pulse arrival -- the other half might be
an interpulse or else unobserved). The general problem is addressed
in Appendix \ref{app:timing}, where the three different types
of solutions are treated separately in Appendices \ref{app:pulsibpos}
and \ref{app:pulsibneg} for positive and negative $I_B$, respectively.
Approximate solutions are also derived in those appendices
(Eqs. [\ref{approxsolnposksmall}], [\ref{approxsolposkbig}]
and [\ref{approxsolneg}]), but those results only apply when
the star is nearly axisymmetric {\it and} $\vert\Omhat_-\vert
\ll 1$. 

Here, we shall investigate two different nearly
axisymmetric cases, $\vert I_B/C\vert\ll 1$ and $\vert I_B/C\vert
\gg 1$. For $\vert I_B\vert\ll C$, it will turn out that
$\vert\Omhat_-\vert\ll 1$, and the results of Appendices
\ref{app:pulsibpos} and \ref{app:pulsibneg} will be directly
applicable. However, for $\vert I_B\vert\gg C$, we have already
mentioned that $\Omhat_-$ need not be small (see, for example,
discussion following Eq. [\ref{omhatpm}]).

\subsection{Pulse Arrival Times for $\vert I_B\vert\ll C$}
\label{pulsibsmall}

When $\vert I_B\vert\ll C$, we can apply the results in
Eqs. (\ref{approxsolnposksmall}) and (\ref{approxsolneg})
directly. Using the approximations given in Appendix
~\ref{app:ibllc}, and accounting for pulsar spindown using
the results of \S~\ref{spindown}, we find that the oscillating
parts of the phase residuals are
\be
\Omega_0\tosc\simeq \Omhat_-\cot\chi\left(-\sin\tau
+{\Omega_0\cos\tau\over t_{sd}\omega_p^2}\right)
\label{oscibposllc}
\ee
for $I_B>0$, and
\be
\Omega_0\tosc\simeq \Omhat_-\cot\chi\left(-\cos\tau
+{\Omega_0\sin\tau\over t_{sd}\omega_p^2}\right)~
\label{oscibnegllc}
\ee
for $I_B<0$, where 
\be
\Omhat_-\simeq\omo_--{\omo_+I_B\sin 2\theta\over 2C}
\label{omo}
\ee
and $\omo_-\equiv\sin(\chi-\theta)$, which is zero in the minimum
energy configuration. (See Appendix \ref{app:ibllc}.) Thus,
to lowest order in the (presumed) small quantity $\Omhat_-\cot\chi$,
phase residuals oscillate only at $\omega_p$. Oscillations at higher
harmonics, such as $2\omega_p$, have amplitudes that are smaller 
by additional factors of $\Omhat_-$. It is possible for these to
be comparable in magnitude to the terms $\sim\Omhat_-$ retained
in Eqs. (\ref{oscibposllc}) and (\ref{oscibnegllc}), but only if
$\cot\chi$ is very small i.e. if $\vert\Omhat_-\cot\chi\vert
\sim\Omhat_-^2$, or $\vert\cot\chi\vert\sim\vert\Omhat_-\vert
\ll 1$. This was also found by Link \& Epstein (2001), who required
$\chi$ very close to $\pi/2$ in order for their axisymmetric precession
model to account for the observed precession of PSR 1828-11. (See also
Rezania 2002.) Here, we
also note that $k^2\simeq e^2\Omhat_-^2$, where $e^2$ represents the
deviation of the star from axisymmetry. Thus, nonaxisymmetric effects
alone cannot introduce substantial harmonic structure in the phase
residuals for small $\Omhat_-^2$ either.

The importance of spindown in the timing residuals is measure
by the nondimensional parameter
\be
\Gamma_{sd}\equiv
{\Omega_0\over\omega_p^2t_{sd}}={P_p^2\over 2\pi P_0t_{sd}}
\simeq 376P_{p,1000}^2P_0^{-1}t_{sd,5}^{-1}
\label{gammasdef}
\ee
where $P_p=1000P_{p,1000}$ days is the precession period,
$P_0$ is the pulsar period in seconds, and
$t_{sd}=10^5t_{sd,5}$ years. For PSR 1828-11, we have
$\Gamma_{sd}\simeq 844$, so the pulsar spindown dominates the 
oscillatory terms. 

\subsection{Pulse Arrival Times for $\vert I_B\vert\gg C$}
\label{pulsiblarge}

When $\vert I_B\vert\gg C$, the moment of inertia tensor is
once again approximately axisymmetric, but neither $\Omhat_-$
nor $\Omhat_+$ has to be small. Thus, the expansions in 
Appendices \ref{app:pulsibpos} and \ref{app:pulsibneg}
are not applicable, and we shall have to solve the timing
equation in a different way. For doing so, Eqs. (\ref{approxibggcpos})
and (\ref{approxibggcneg}) prove to be useful. To keep the
notation compact, we define the nondimensional parameter
$\Ctil\equiv C/\vert I_B\vert$.

Let us consider the two possible sign choices separately. For
$I_B>0$, we use Eq. (\ref{bhatykbig}) to evaluate $\bhat_y$.
The problem is simplified since, from Eq. (\ref{approxibggcpos}),
$\bhat_+\simeq\Ctil\sin 2\theta/2\ll 1$. Consequently,
to first order in $\Ctil$, we find
\be
\bhat_y\simeq\sin\chi\cos\phi-{\Ctil\sin 2\theta\over 2}
\left(\sin\tau\sin\phi+\cos\chi\cos\phi\right)~,
\ee
so pulses arrive when 
\be
\phi\simeq\left(2n+{1\over 2}\right)\pi-{\Ctil\sin 2\theta
\sin\tautil\over 2\sin\chi}~;
\ee
mapping from phase to time implies
\be
\int_0^t{dt~L\over\Ibar}\simeq\left(2n+{1\over 2}\right)\pi-{\Ctil\sin 2\theta 
\sin\tautil\over 2\sin\chi}
+{\Ctil\sin^2\theta\sin 2\tautil\over 4\cos\chi}~.
\ee
Taking account of pulsar spindown we find that the oscillatory part
of the timing residuals is
\be
\Omega_0\tosc\simeq\Ctil\left[{\sin^2\theta\sin 2\tautil\over 4\cos\chi}
-{\sin 2\theta\sin\tautil\over 2\sin\chi}
+\Gamma_{sd}\left({\sin^2\theta\cos 2\tautil\over 16}
+{\cos\chi\sin 2\theta\cos\tautil\over 2\sin\chi}\right)\right]~.
\label{oscibggcpos}
\ee
Note that the phase residuals vanish as $\Ctil\to 0$ even though
the star still precesses. Moreover, there are oscillations at
both $\omega_p$ and $2\omega_p$ whose amplitudes may be comparable,
in agreement with observations of PSR 1828-11.

For $I_B<0$, we find, to order $\Ctil$, we find that
pulses arrive when
\be
\int{dt~L\over\Ibar}\simeq\left(2n+{1\over 2}\right)\pi
-{\Ctil\sin 2\theta\cos\tau\over 2\sin\chi}
-{\Ctil\sin^2\theta\sin 2\tau\over 4\cos\chi}~,
\ee
and taking account of spindown results in oscilating
timing residuals
\be
\Omega_0\tosc\simeq
\Ctil\left[-{\sin 2\theta\cos\tau\over 2\sin\chi}
-{\sin^2\theta\sin 2\tau\over 4\cos\chi}
+\Gamma_{sd}\left({\sin^2\theta\cos 2\tau\over 16}
-{\sin 2\theta\cos\chi\sin\tau\over 2\sin\chi}\right)
\right]~.
\label{oscibggcneg}
\ee
Once again, this involves oscillations at both $\omega_p$
and $2\omega_p$ which can have comparable magnitudes.
We see again that as $\Ctil\to 0$, the oscillatory phase
residuals disappear.

\subsection{Application to PSR 1828-11}
\label{application}

For PSR 1828-11, timing residuals appear to oscillate at both
$\omega_p$ and $2\omega_p$ with similar amplitude. The results
of \S \ref{pulsibsmall} show that this situation is incompatible
with small values of $\vert\Omhat_-\vert$ if $\vert I_B\vert\ll C$.
Moreover, we note that the results of \S \ref{pulsibsmall} continue
to hold as $\vert I_B\vert\to 0$, so we see that equal amplitudes
at $\omega_p$ and $2\omega_p$ cannot arise from a model without
magnetic stresses, but with a triaxial crust, unless there is
some fine-tuning of parameters (as in the axisymmetric model
of Link \& Epstein [2001], which requires $\chi$ very close
to $\pi/2$).

Thus, we focus on the strongly magnetic case, $\vert I_B\vert
\gg C$. In this case, Eqs. (\ref{oscibggcpos}) and
(\ref{oscibggcneg}) show that it is possible for the phase
residuals to oscillate with comparable amplitudes.  The difference
between the large and small $\vert I_B\vert/C$ limits is that
for small $\vert I_B\vert/C$, the amplitude of the observed
timing residuals is determined
solely by $\vert\Omhat_-\vert$, but at large  $\vert I_B\vert/C$
the amplitude is determined by $C/\vert I_B\vert$
primarily.  Thus, by contrast to what we found for $\vert I_B\vert
\ll C$, small amplitude need not suppress the 
oscillations at $2\omega_p$.   

For PSR 1828-11, we also know that $\Gamma_{sd}\simeq 844\gg 1$, so
let us approximate the oscillatory timing residuals further as
\baray
\Omega_0\tosc&\simeq&{\Ctil\Gamma_{sd}\sin^2\theta\over 16}\left(
\cos 2\tautil+{16\cos\tautil\over\tan\theta\tan\chi}\right)
\qquad\qquad(I_B>0)
\nonumber\\
\Omega_0\tosc&\simeq&{\Ctil\Gamma_{sd}\sin^2\theta\over 16}\left(
\cos 2\tau-{16\sin\tau\over\tan\chi\tan\theta}\right)
\qquad\qquad(I_B<0)~.
\label{largespindown}
\earay
The parameter
\be
u\equiv{16\over\tan\chi\tan\theta}
\label{udef}
\ee
governs the relative strengths of the two harmonics in the timing
residuals. If the precession is in the minimum energy state,
$\theta=\chi$, then $u=1$ for $\chi\simeq 76^\circ$. 

The timing residuals in Eq. (\ref{largespindown}) vary between
different minimum and maximum values. For $I_B>0$, $\Omega_0
\tosc$ is minimum when $\cos\tautil=-u/4$, provided that
$\vert\tan\chi\tan\theta\vert<1$; for $\chi=\theta$, this is
so as long as $\chi>63^\circ$. Presuming this to be so,
the minimum value is 
\be
(\Omega_0\tosc)_{min}=-{\Ctil\Gamma_{sd}\sin^2\theta\over 16}
\left(1+{u^2\over 8}\right)~.
\label{ibposmin}
\ee
The maximum is at $\cos\tautil=1$, where we find
\be
(\Omega_0\tosc)_{max}={\Ctil\Gamma_{sd}\sin^2\theta\over 16}
\left(1+u\right)~.
\label{ibposmax}
\ee
The ratio of maximum to minimum timing residual is
\be
{(\Omega_0\tosc)_{max}\over\vert(\Omega_0\tosc)_{min}}\vert
={1+u\over 1+u^2/8}\qquad\qquad(I_B>0)~;
\ee
for $u=1$, this ratio is $16/9\simeq 1.8$, and the ratio is
two for $u=2$. Observationally, the timing residuals for
PSR 1828-11 appear skewed toward positive values, with a
maximum about twice the magnitude of the minimum (see Fig.
1 in Stairs, Lyne \& Shemar [2000]). Similarly, for $I_B<0$,
the maximum value of $\Omega_0\tosc$ occurs when $\sin\tau
=-u/4$, and the minimum occurs when $\sin\tau=1$; in this
case, the ratio of minimum to maximum values is
\be
{(\Omega_0\tosc)_{max}\over\vert(\Omega_0\tosc)_{min}\vert}
={1+u^2/8\over 1+u}\qquad\qquad(I_B<0)~;
\ee
for $u=1$, this ratio is $9/16\simeq 0.56$, and the ratio
is two for either $u\simeq 16.9$ or $u\simeq -0.89$. To the
extent that we expect $u>0$ (manifestly so for $\chi=\theta$),
and $u\sim 1$ (but not $\gg 1$)
the observed residual arrival times
appear to favor $I_B>0$, i.e. {\it prolate}
magnetic distortions.

\begin{figure*}
\centerline{\psfig{figure=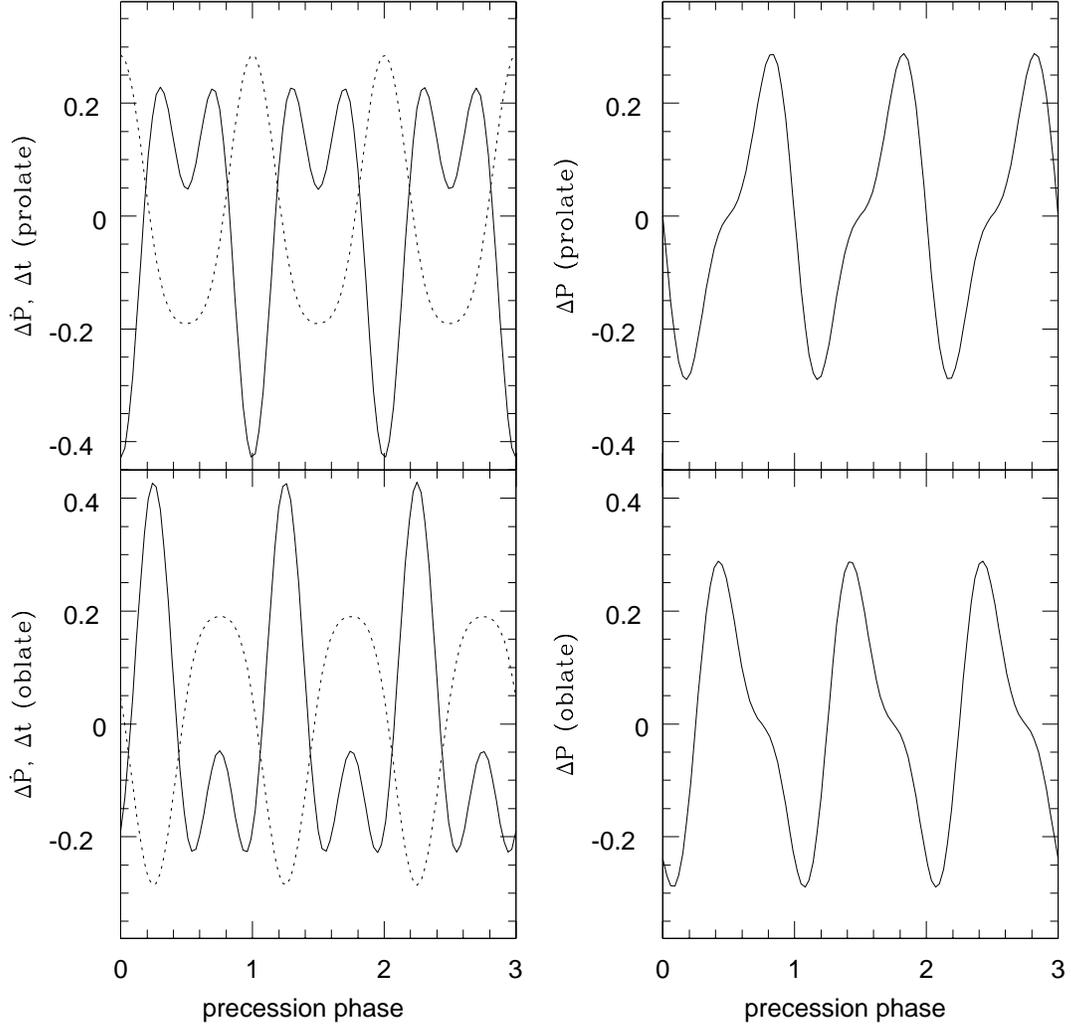,width=0.9\hsize}}
\caption[]{
Results of evaluating the oscillating residual arrival
time $\Delta t$ and its first two derivatives
$\Delta P$ and $\Delta\dot P$ for $u=5$ and $\chi=\theta$.
The top panels are for prolate models, the bottom for
oblate.  The left panels show $\Delta t$ (dotted) and
$\Delta\dot P$ (solid), and the right panels shwo
$\Delta P$. The units of $\Delta t$ are $\Ctil\Gamma/\Omega_0$,
the units of $\Delta P$ are $2\pi\Ctil\Gamma\omega_p/\Omega_0^2
=\Ctil\Gamma P_0^2/P_p$ (approximately $1.60\Ctil\mu{\rm s}$
for PSR 1828-11),
and the units of $\Delta\dot P$ are $2\pi\Ctil\Gamma\omega_p^2/
\Omega_0^2=2\pi\Ctil\Gamma P_0^2/P_p^2$ (approximately $1.17
\Ctil\times 10^{-13}$ for PSR 1828-11).
}
\label{fig:residualexample}
\end{figure*}

Fig. \ref{fig:residualexample}
illustrates the residuals in the arrival times, $\Delta t$
(dotted, left panel), period derivatives $\Delta\dot P$ 
(solid, dotted) and period $\Delta P$ (solid, right panel)
for $\theta=\chi$ and $u=5$ for the prolate (upper) and
oblate (lower) cases. The period an period derivatives
are computed by differentiating $\Omega_0\tosc$ with respect
to time:
\baray
\Delta P(t)&=&{2\pi\over\Omega_0^2}{d(\Omega_0\tosc)\over dt}
={P_0^2\over P_p}{d(\Omega_0\tosc)\over d\tau}
\nonumber\\
\Delta\dot P(t)&=&{2\pi\over\Omega_0^2}
{d^2(\Omega_0\tosc)\over dt^2}=
2\pi\left({P_0\over P_p}\right)^2{d^2(\Omega_0\tosc)\over d
\tau^2}~.
\earay
The agreement between these evaluations
and the results plotted in Fig. 2 of Stairs, Lyne \& Shemar
(2000) is good, superficially, for the prolate model. Better
agreement is seen for the variable period and period derivative
than for the arrival times themselves; this may have been
expected (e.g. Cordes 1993).

Using the results plotted in Fig. \ref{fig:residualexample}, 
we can estimate the
magnitude of $\Ctil$ and the angles involved, even though
these results do not constitute a true fit to the data,
but just a plausible model. The curves for $\Delta\dot P$
and $\Delta P$ resemble the observational results better,
so let us focus on those. 
Fig. \ref{fig:residualexample}
was prepared for $\chi=\theta$ and $u=5$, which corresponds
to $\chi=\theta=60.8^\circ$. From Fig. 2 in Stairs, Lyne
\& Shemar (2000), we see that the maximum values of
$\Delta P$ and $\Delta\dot P$ are about 1 ns
and $0.2\times 10^{-15}$ for PSR 1828-11. For the prolate
model, which resembles the observations better,
Fig. \ref{fig:residualexample}
implies a peak value of $460\Ctil$ ns for $\Delta P$,
and $2.6\Ctil\times 10^{-14}$ for $\Delta\dot P$.
We therefore estimate $\Ctil\simeq
0.002$ from $\Delta P$, and 
$\Ctil\simeq 0.008$ from $\Delta\dot P$. Since we have
not attempted true curve fits (i.e. by varying the parameters
$u$, $\theta$ and $\Ctil$)
we regard this as acceptable agreement, provisionally.

\section{Discussion}
\label{discussion}

Here, we have extended previous studies of precession of neutron 
stars to incorporate the effects of oblique magnetic fields. We have
shown that if the magnetic stresses are large enough, then steady
rotation is unlikely, and the neutron star must precess. Moreover,
even when the magnetic stresses are relatively weak, so that steady
rotation is possible irrespective of the obliquity of the magnetic
field, the minimum energy state is not the steady state. Thus,
even in this case, the neutron star will precess.
We argued, in \S \ref{precess}, that the minimum energy,
precessing state is a local energy minimum that applies at fixed
angle between the magnetic and rotational axes. On a longer
timescale, we would expect the star to seek its global energy
minimum, which should correspond to either aligned or perpendicular
magnetic and rotation axes, and no precession. We might expect
short timescale dissipative effects to drive the system toward
its local minimum, and that the global minimum is only achieved
on somewhat longer timescale, perhaps as a result of electromagnetic
spindown torques (e.g. Goldreich 1970).

The effective moment of inertia tensor of a neutron star with an
inclined magnetic field is inherently triaxial. Consequently, the
precession is {\it periodic} but not {\it sinusoidal} in time.
In general, the solution for the rotational angular velocity of the
star can be expanded in a Fourier series involving harmonics of
the precession frequency. We have shown that at least the first
few terms in such an expansion can have comparable magnitudes 
provided that the interior magnetic stresses are not very small.

The condition that magnetic stresses play an important role
is that the magnetic-induced distortions are comparable to
or larger than the
distortions of the stellar crust. For precession periods of
order years, the implied magnetic stresses exceed those expected
from the classical Maxwell stress tensor, evaluated using the
inferred dipole magnetic field strength, by a couple of orders
of magnitude. However, if the interior of a neutron star contains
a Type II superconductor, or else is a normal conductor
but possesses large toroidal
magnetic fields, the magnetic stresses are larger,
and the implied distortions can be of the right order of magnitude
(Jones 1975, Easson \& Pethick 1977, Cutler 2002).
Thus, the observation of neutron star precession can be taken
as indirect evidence for enhanced magnetic stresses, due to either
Type II superconductivity, or large toroidal fields.

We postpone detailed application of the ideas set forth here
to PSR 1828-11 to another paper (Akgun, Epstein \& Wasserman
2002). However, in \S \ref{application} we argued that 
only a model with $\vert I_B\vert\gg C$ can lead to time
residuals that oscillate with comparable amplitude at both
$\omega_p$ and $2\omega_p$. In this case, the amplitude
of the observed time residuals is set by the dimensionless
ratio $\Ctil\equiv C/\vert I_B\vert$, not by the tilt of the angular
velocity vector away from any principal axis, which may
be large.

By contrast, if the stellar distortions associated
with magnetic stresses are {\it smaller than} those
associated with the crust, then Eqs. (\ref{oscibposllc})
and (\ref{oscibnegllc})
show that the timing residuals oscillate predominantly
at $\omega_p$. Oscillations at $2\omega_p$ would be
down by factors $\sim\vert\Omhat_-\vert$, where
$\Omhat_-$ is given by Eq. (\ref{omo}), and is
$\sim\vert I_B\vert/C$ in the minimum energy configuration.
Moreover, we argued, in \S \ref{pulsibsmall} and
\S \ref{application}, that
precession of a triaxial crust alone would probably
not, at small precession amplitude, be capable of producing
oscillations of comparable magnitude at both $\omega_p$
and $2\omega_p$, because the precession amplitude is
proportional to $\vert\Omhat_-\vert$ and the oscillations at
harmonics of $\omega_p$ are suppressed by factors
$\sim\vert\Omhat_-\vert$. Thus, a solution
in which the crustal distortion is responsible for precession
is unlikely to explain the data on PSR 1828-11.
The model in which magnetic stresses dominates is not merely
precession of a triaxial body because small amplitude phase
residuals can arise even when $\vert\Omhat_-\vert$ is
not small.

For a precessing star in which the distortion is due to
magnetic stresses primarily, the precession period is
%(see Eq. (\ref{ompibigneg})
\be
{P_p}\simeq{P_0\Ibar\over 3I_B\cos\chi}
\simeq{492~{\rm days}\over\beta\cos\chi(BH)_{27}/5}~,
\ee
where the value given is for PSR 1828-11. This is
similar to the observed period,
about 1000 days, provided that $\beta\cos\chi$ is not
very small. Thus, if $\chi\sim 60^\circ$
the timing residuals could oscillate with about the 
right period, amplitude and relative importance of the
fundamental precession period and its first harmonic.
In this case, we note that the condition that there is
no steady state, $\vert\sin 2\chi\vert\geq\Ctil$ can
be satisfied: for $\chi=60^\circ$, for example,
$\sin 2\chi=\sqrt{3}/2$ whereas we estimated that
$\Ctil\sim 0.001-0.01$ in \S \ref{application}. This
possible explanation of the timing residuals for
PSR 1828-11 only works if the magnetic stresses
in this star are $\sim 200$ times larger than would be indicated
by its dipole magnetic field. Thus, we conclude that
either the interior is a Type II superconductor,
or is a normal conductor with 
a toroidal field whose strength is
$\sim 10^{14}$ G. Otherwise, the expected magnetic stresses
are far smaller than is needed for this solution to apply.

We also noted, in \S \ref{application}, that the 
ratio of the minimum and maximum timing residuals,
and the shape of the variation of $\Delta P$ and
$\Delta\dot P$ 
seen in PSR 1828-11 appear to favor a model in which
$I_B>0$, so that magnetic distortions are {\it prolate}.
Prolate distortions would arise naturally 
from the stresses due to a toroidal
field, with or without Type II superconductivity
(Cutler 2002), but may also result if magnetic flux
tubes have been transported outward in the core
and accumulate at its outer boundary
(Ruderman, Zhu \& Chen 1998, Ruderman \& Chen 1999), 
which could ``pinch'' the interior.

Crustal distortions are still needed in order for the precessional
amplitude to be nonzero. In fact, to be more precise, the 
precessional amplitude depends on the component of the moment
of inertia tensor of the crust that is {\it not} symmetric about
the magnetic axis. This can be seen directly from Eqs. (\ref{oscibggcpos})
and (\ref{oscibggcneg}), which show that the oscillating
time residuals vanish as $\Ctil\sin\theta\to 0$, where $\theta$
is the angle between the magnetic axis $\bhat$ and, in this axisymmetric
distortion model, the symmetry axis of the relevant crustal
deformation, $\khat$.
Thus, although all neutron stars
precess as a consequence of their magnetic stresses
for $\vert I_B\vert\gg C$ in the
picture advanced here, only those with sufficiently large
nonaligned
crustal deformations would have discernible oscillations of
their timing residuals. In this sense, PSR 1828-11 may be special.

Although we have treated the basic physics of precession of an
oblique rotator in some detail, we have not treated several
effects that might play significant roles. We have not
explicitly included either vortex line pinning or vortex drag.
Link \& Cutler (2002) have argued that vortex lines can unpin
globally at large enough precession amplitude. For $\vert I_B
\vert\gg C$, as is required to explain the timing residuals
in PSR 1828-11, the precession amplitude is $\simeq\sin\chi$
or $\cos\chi$ (depending on the sign of $I_B$),
which is not small, so global unpinning is expected. Thus,
we may expect that vortex lines are unpinned in neutron stars
with magnetic fields that are strong enough to have $\vert I_B
\vert\gg C$ except for $\chi$ very close to either zero or
$\pi/2$, depending on the sign of $I_B$. For these, vortex line
drag, if weak enough, may simply serve to bring the neutron
star superfluid into corotation with its crust, and drive
the rotating star toward its minimum energy state. For 
large $\vert I_B\vert/C$, we have seen that precession
is required, so weak vortex drag or other
forms of dissipation need not prevent precession.
For small values of $\vert I_B\vert/C$,
the precession amplitude is given by Eq. (\ref{omo}) and can
be very small; in the minimum energy state, Eq. (\ref{omo})
implies $\vert\Omhat_-\vert\simeq\vert I_B\sin 2\theta\vert/C$.
In this case, it is possible that precession cannot overcome
pinning forces, as discussed in Link \& Cutler (2002), and so
precession does not occur.

Theories of pulsar glitches involve the pinning, unpinning and
repinning of crustal superfluid vortex lines
(e.g. Anderson and Itoh 1975; Alpar et al. 1981, 1984a,b, 1993; Link,
Epstein and Baym 1993).
As we have seen, for large $\vert I_B\vert/C$, precession amplitudes
are large, and vortices may be expected to unpin, 
but it is possible for vortex lines
to remain pinned in neutron stars with $C\gg\vert I_B\vert$.
Thus, there could be a dichotomy between pulsars that
glitch ($C\gg\vert I_B\vert$) and those that precess
($C\ll\vert I_B\vert$). If, in the course of a glitch, all
crustal superfluid vortices were to unpin, then the star
might precess briefly. Perhaps that explains the detection of
damped, quasisinusoidal timing residuals in the Vela pulsar after
and perhaps before its Christmas 1988 glitch (McCulloch
et al. 1990).

We have kept the problem of precession of an oblique rotator
as simple as possible by considering what happens when the
magnetic field is axisymmetric about some axis, and the
crustal distortions are also axisymmetric, but about a
different axis. More realistically, both of these simplifying
assumptions are likely to be violated. Most likely, the
crust is not axisymmetric. When magnetic stresses dominate,
we do not expect including intrinsic crustal asymmetry to
alter the results found here qualitatively, since the effective
moment of inertia is already triaxial here. Triaxiality of
the crust, in the limit of rather small crustal distortions,
would simply rotate the principal axes slightly. 
%In the limit
%where crustal distortions dominate, though, there could be 
%more significant changes in our results, particularly the
%conclusion that, in that limit, timing residuals oscillate
%predominantly at $\omega_p$, and only slightly at $2\omega_p$.
Furthermore, the magnetic field may have a more complicated
structure than we have assumed. A substantial quadrupolar
component would presumably render the contribution to the
inertia tensor from magnetic stresses alone triaxial. We
shall consider these complications elsewhere.

Although we have included the spindown torque in our evaluations
of timing residuals, we did not include near zone electromagnetic
torques (Good \& Ng 1985, Melatos 1997, 1999, 2000).
The principal effect of such torques would be to renormalize
the moment of inertia tensor of the star.
Near zone torques can play a role similar
to the magnetic distortions considered here, but are smaller
by a factor $\sim (H/B)(Rc^2/GM)$, which is non-negligible
even if $H=B$. However, we note here that the large magnetic
distortions we propose would presumably apply to 
the spindown of magnetars
and anomalous X-ray pulsars, in much the same fashion as
proposed by Melatos (1999, 2000). We shall pursue this
idea elsewhere.

We have also ignored motions of the fluid and crust of the
star apart from rigid rotation.
Mestel and collaborators (Mestel \& Takhar 1972, Mestel et al. 1981, 
Nittman \& Wood 1981)
have pointed out that the equilibrium in a fluid star with
oblique magnetic field must involve fluid motions 
with velocities $\sim(\Omega^2R^3/GM)\omega_pR$. These
distort the stellar magnetic field by a fractional amount
$\sim \Omega^2R^3/GM$ over the precession period $2\pi/\omega_p$
(Mestel \& Takhar 1972, Mestel et al. 1981, Nittman \& Wood 1981).
Similar motions
might arise in the crust, but with magnitudes 
$\sim(\Omega^2R^2/c_t^2)\omega_pR$, where $c_t$ is the
sound speed for transverse waves. The expected amplitude
of the resulting magnetic wander is $\sim \Omega^2R^2/c_t^2$.
Although slow, these
displacements cause the magnetic field of the star to
oscillate about its undisturbed, axisymmetric state, and might
influence the long-term behavior of the observed spindown. We
shall investigate whether there is any long-term observational
signature of these motions elsewhere.
In addition, it is likely that magnetic and rotational 
deformations of the core must also deform the crust, as they
exert pressure on its inner boundary. We would expect the magnetic
deformation of the core to promote crustal deformation symmetric
about $\bhat$, which would not lead to observable precession, but
the rotation-induced deformation need not be symmetric about
$\bhat$, and should be substantial. An important question left
unanswered here is whether the crustal distortion, $C_{ij}$, that leads
to detectable precession is relatively steady, or is simply due
to a seismic fluctuation. We shall explore the important issue of 
crustal deformations elsewhere.

Finally, we emphasize that any neutron star with strong enough
core magnetic stresses ought to precess, but we may
not be able to detect their precession because their magnetic axes can still
rotate more or less uniformly. This is because, at small values
of $C/\vert I_B\vert$, the neutron star precesses almost exactly
about its magnetic axis, which therefore rotates almost uniformly
as seen in the inertial frame. 
%PSR 1828-11 may be special, possessing
%crustal distortions that are larger than typical for some reason.
Although the precession may not be detectable readily from 
timing residuals for most pulsars, gravitational radiation
amplitudes would be larger than would arise
without enhanced internal magnetic stresses (e.g. Cutler \& Thorne 
2002, Cutler 2002). The distortions required for PSR 1828-11 are still smaller
than would be needed for detection by LIGO, even if it were spinning
faster (Brady et al. 1998).
If there are young, highly magnetized neutron stars rotating rapidly,
they would be the brightest emitters of gravitational radiation.
Such objects have been hypothesized to be the sources of the
highest energy cosmic rays (Blasi, Epstein \& Olinto 2000,
Arons 2002).

\acknowledgments

Partial support for this work was provided by a grant from
IGPP at LANL. I thank T. Akgun, J. Cordes, R. Epstein and
B. Link for comments.

\appendix
\section{Inequalities Among Eigenvalues}
\label{app:appendixa}

For $I_B>0$, we can rewrite Eq. (\ref{eigenvalues}) in the form
\be
\delta I_\pm={3\over 2}\left[C-I_B
\pm\sqrt{(C-I_B)^2+2CI_B(1-\cos 2\theta)}\right]~,
\label{irewritepos}
\ee
from which it follows that
\baray
\delta I_+&\geq&{3\over 2}\left(C-I_B+\vert C-I_B\vert\right)
\nonumber\\
\delta I_-&\leq&{3\over 2}\left(C-I_B-\vert C-I_B\vert\right)~.
\label{inequalpos}
\earay
Eq. (\ref{inequalpos}) implies that $\delta I_+\geq 3(C-I_B)>0$ and
$\delta I_-<0$ if $I_B<C$, and $\delta I_+>0$ and $\delta I_-<
3(C-I_B)<0$ if $I_B>C$. Thus, if $I_B>0$, then $\delta I_+>0$ and
$\delta I_-<0$ irrespective of whether $I_B>C$ or $I_B<C$.
Thus, for $I_B>0$, $I_+>I_2>I_-$.
 
For $I_B<0$, we can rewrite Eq. (\ref{eigenvalues}) in the form
\be
\delta I_\pm={3\over 2}\left[C+\vert I_B\vert
\pm\sqrt{(C-\vert I_B\vert)^2+2C\vert I_B\vert(1+\cos 2\theta)}\right]~,
\label{irewriteneg}
\ee
from which it follows that
\baray
\delta I_+&\geq&{3\over 2}\left(C+\vert I_B\vert
+\vert C-\vert I_B\vert\vert\right)\nonumber\\
\delta I_-&\geq&{3\over 2}\left(C+\vert I_B\vert
-\vert C+\vert I_B\vert\vert\right)=0~.
\label{inequalneg}
\earay
Eq. (\ref{inequalneg}) implies that $\delta I_+>3C$ if
$\vert I_B\vert<C$ and $\delta I_+>3\vert I_B\vert$
if $\vert I_B\vert>C$. Thus, we see that for $I_B<0$,
$I_+>I_->I_2$.

\section{Approximate Results for $\vert I_B\vert\ll C$}
\label{app:ibllc}  

For $\vert I_B\vert\ll C$, we can expand these results to find
\baray   
%{1\over\sqrt{2}}\left[1+{\eta\over\sqrt{1+\eta^2}}\right^{1/2}
\sin\sigma_+=-\bhat_-
&\simeq& \sin\theta\left[1+{I_B\cos^2\theta\over C}
+{I_B^2(5\cos^4\theta-3\cos^2\theta)\over 2C^2}+\cdots\right]
\nonumber\\
%{1\over\sqrt{2}}\left[1-{\eta\over\sqrt{1+\eta^2}}\right^{1/2}
\cos\sigma_+=\bhat_+
&\simeq& \cos\theta\left[1-{I_B\sin^2\theta\over C}
+{I_B^2(5\sin^4\theta-3\sin^2\theta)\over 2C^2}+\cdots\right]
\nonumber\\
\delta I_+ &\simeq& 3C\left[1-{I_B(1+\cos 2\theta)\over 2C}
+{I_B^2\sin^22\theta\over 4C^2}+\cdots\right]
\nonumber\\
\delta I_- &\simeq& -{3I_B(1-\cos 2\theta)\over 2}
-{3I_B^2\sin^22\theta\over 4C^2}+\cdots~.
\label{smallib}
\earay   
Using Eq. (\ref{smallib}) we also find that
\baray
\Omhat_+&\simeq&\omo_++{\omo_-I_B\sin 2\theta\over 2C}+\cdots
\nonumber\\
\Omhat_-&\simeq &\omo_--{\omo_+I_B\sin 2\theta\over 2C}+\cdots
\nonumber\\
\Delta&\simeq&{3C\over 2\Ibar}\left(1-{I_B\cos 2\theta\over C}+\cdots\right)
\nonumber\\
\Delta_0&=&{3C\over 2\Ibar}\left(1-{I_B\over C}\right)~,
\label{smallibstuff}
\earay
where $\omo_+\equiv\cos(\chi-\theta)$ and $\omo_-\equiv\sin(\chi-\theta)$.
%
%\be
%\sigma_+\simeq\theta+{I_B\sin 2\theta\over 2C}+{I_B^2\sin 2\theta
%\cos 2\theta\over 2C^2}+\cdots~,
%\ee
%and $\sigma_-=\sigma_++\pi/2$. As expected, $\ehatplus$ is nearly
%along $\khat$ in this case, and the largest eigenvalue is nearly
%$I_2+3C$. For $I_B>0$, the intermediate eigenvalue is slightly
%smaller than $I_2$, the smallest eigenvalue, and the associated
%eigenvector is in the $\khat-\bhat$ plane, at an angle $\pi/2$
%relative to $\ehatplus$. For $I_B<0$, the smallest eigenvalue
%is $I_2$, which is only slighly smaller than the intermediate
%eigenvalue, whose eigenvector is in the $\khat-\bhat$ plane, at an angle $\pi/2$
%relative to $\ehatplus$.

\section{Approximate Results for $\vert I_B\vert\gg C$}
\label{app:ibggc}

For $\vert I_B\vert\gg C$, we find
\baray
1\pm{\eta\over\sqrt{1+\eta^2}}&\simeq& 1\pm s_B
\left[1-{C^2\sin^22\theta\over 2I_B^2}+\cdots\right]
\nonumber\\
\delta I_+&\simeq&{3\vert I_B\vert\over 2}\left[1-s_B+{C(1-s_B\cos 2\theta)
\over\vert I_B\vert}
+{C^2\sin^22\theta\over 2I_B^2}+\cdots\right]\nonumber\\
\nonumber\\
\delta I_-&\simeq&{3\vert I_B\vert\over 2}\left[-1-s_B+{C(1+s_B\cos 2\theta)
\over\vert I_B\vert}-{C^2\sin^22\theta\over 2I_B^2}+\cdots\right]~.
\earay
where $s_B\equiv I_B/\vert I_B\vert$ is the sign of $I_B$.
For $I_B>0$ these results imply
\baray
\sin\sigma_+=-\bhat_-&\simeq& 1-{C^2\sin^22\theta\over 8}+\cdots
\nonumber\\
\cos\sigma_+=\bhat_+&\simeq&{C\sin 2\theta\over 2I_B}+\cdots
\nonumber\\
\Omhat_+&\simeq&\sin\chi+{C\cos\chi\sin 2\theta\over 2I_B}+\cdots
\nonumber\\
\Omhat_-&\simeq& -\cos\chi+{C\sin\chi\sin 2\theta\over 2I_B}+\cdots
\nonumber\\
\Delta&\simeq&{3I_B\over 2\Ibar}\left(1-{C\cos 2\theta\over I_B}
+\cdots\right)
\nonumber\\
\etil^2&\simeq&{\Delta+\Delta_0\over\Delta-\Delta_0}\simeq
{C\sin^2\theta\over I_B}
\nonumber\\
\omega_p&\simeq& {2\Delta L\cos\chi\over\Ibar}~,
\label{approxibggcpos}
\earay
and for $I_B<0$ the same results imply
\baray
\sin\sigma_+=-\bhat_-&\simeq&{C\sin 2\theta\over 2\vert I_B\vert}+\cdots
\nonumber\\
\cos\sigma_+=\bhat_+&\simeq& 1-{C^2\sin^22\theta\over *I_B^2}+\cdots
\nonumber\\
\Omhat_+&\simeq&\cos\chi+{C\sin\chi\sin 2\theta\over 2\vert I_B\vert}+\cdots
\nonumber\\
\Omhat_-&\simeq&\sin\chi-{C\cos\chi\sin 2\theta\over 2\vert I_B\vert}+\cdots
\nonumber\\
\Delta&\simeq&{3\vert I_B\vert\over 2\Ibar}\left(1+{C\cos 2\theta\over
\vert I_B\vert}+\cdots\right)
\nonumber\\
e^2&\simeq&{\Delta_0-\Delta\over 2\Delta}\simeq{C\sin^2\theta\over
\vert I_B\vert}+\cdots
\ nonumber\\
\omega_p&\simeq&{2\Delta L\sqrt{e^2+\Omhat_+^2}\over I}
\simeq {2\Delta L\cos\chi\over \Ibar}~.
\label{approxibggcneg}
\earay
%Therefore, we find that
%\be
%\sigma_+\simeq {\pi\over 2}-{C\vert\sin 2\theta\vert\over 2I_B}
%\ee
%if $s_B=+1$, and
%\be
%\sigma_+\simeq {C^2\sin^22\theta\over 2I_B^2}
%\ee
%if $s_B=-1$; in both cases, $\sigma_-=\sigma_++\pi/2$.
%Thus, if $I_B>0$, the principal axis of largest moment of inertia is
%approximately perpendicular to $\bhat$, and has an eigenvalue only
%slightly larger than $I_2$, whereas the principal axis with smallest
%moment of inertia is nearly along $-\bhat$, and has an eigenvalue
%that is approximately $I_2-3I_B$. If $I_B<0$, then the principal axis
%of largest moment of inertia is approximately along $\bhat$, and
%has an eigenvalue that is approximately $I_2+3\vert I_B\vert$. The
%smallest eigenvalue is $I_2$, and its eigenvector is along $\ehattwo$.
%The intermediate eigenvalue is only slightly larger than $I_2$.
 
\section{Timing Solution}
\label{app:timing}

To find pulse arrival times, we need to determine the motion of
$\bhat$ in the inertial reference frame of the observer.
To do this, we need the Euler angle rotation from the rotating
frame of reference to the inertial frame; for an arbitrary vector
$\Vvec$ this is
(see e.g. Goldstein Eq. [4-47])
\baray
\Vvec_x&=&\Vvec_{\onehat}(\cos\psi\cos\phi-\cos\alpha\sin\psi\sin\phi)
-\Vvec_{\twohat}(\sin\psi\cos\phi+\cos\alpha\cos\psi\sin\phi)
+\Vvec_{\threehat}\sin\alpha\sin\phi\nonumber\\
\Vvec_y&=&\Vvec_{\onehat}(\cos\psi\sin\phi+\cos\alpha\sin\psi\cos\phi)
-\Vvec_{\twohat}(\sin\psi\sin\phi-\cos\alpha\cos\psi\cos\phi)
-\Vvec_{\threehat}\sin\alpha\cos\phi\nonumber\\
\Vvec_z&=&\Vvec_{\onehat}\sin\alpha\sin\psi
+\Vvec_{\twohat}\sin\alpha\cos\psi+\Vvec_{\threehat}\cos\alpha~,
\label{eulerangles}
\earay
where $\alpha,\phi,\psi$ are the Euler angles defined in Fig. 47
of Landau \& Lifshitz, \S 35, except that, to avoid confusion with
our definition of $\theta$ as the angle between $\khat$ and $\bhat$, we
label their Euler angle $\theta$ as $\alpha$. We assume that
$\Lvec$ is along the $\ehatz$ direction in the inertial frame.
We can then determine the two angles $\alpha$ and $\psi$ from the
equations
\baray
L_\threehat&=&L\cos\alpha\nonumber\\
L_\onehat&=&L\sin\alpha\sin\psi\nonumber\\
L_\twohat&=&L\sin\alpha\cos\psi~;
\earay
the third Euler angle $\phi$ is not determined by these relations, but
can be found from
\be
{d\phi\over dt}=L\left({I_\onehat\Omega_\onehat^2+I_\twohat\Omega_\twohat^2
\over I_\onehat^2\Omega_\onehat^2+I_\twohat^2\Omega_\twohat^2}\right)~.
\ee
The choice of axes $\ehat_{\onehat,\twohat,\threehat}$
in the rotating frame of reference is somewhat
arbitrary, and we shall make three different choices below, as the
situation demands.

\subsection{Pulse Arrival Times for $I_B>0$}
\label{app:pulsibpos}

\subsubsection{$I_B>0$ and $k^2<1$}
\label{app:pulsibposksmall}

For $I_B>0$ and $k^2<1$, we choose $\Vvec_{\onehat}=\Vvec_-$, 
$\Vvec_{\twohat}=\Vvec_2$, and $\Vvec_{\threehat}=\Vvec_+$; 
then Eqs. (\ref{eulerangles}) imply (Landau \& Lifshitz Eq. [37.15])
\baray
\cos\alpha&=&{\Omhat_+(1+\Delta)~\dn(\tau)\over
\sqrt{1+2\Delta(2\Omhat_+^2-1)+\Delta^2}}\simeq\Omhat_+
~\dn(\tau)\nonumber\\
\tan\psi&=&{\cn(\tau)\over\sn(\tau)}~
\sqrt{(\Delta+\Delta_0)(1-\Delta)\over 2\Delta(1-\Delta_0)}
\simeq{\cn(\tau)\over\sn(\tau)}~
\sqrt{(\Delta+\Delta_0)\over 2\Delta}~,
\label{alphapsipos}
\earay
where the approximations are to leading order in $\Delta$.
The remaining Euler angle $\phi$ evolves according to
(Landau \& Lifshitz, Eq. [37.16])
\baray
{d\phi\over dt}&=&{L\over\Ibar}\left[{\Delta+\Delta_0+(\Delta-\Delta_0)~\sn^2(\tau)
\over (1-\Delta)(\Delta+\Delta_0)+(\Delta-\Delta_0)(1+\Delta)~\sn^2(\tau)}
\right]\nonumber\\
&\simeq&{L\over\Ibar}\left[1+{\Delta[\Delta+\Delta_0-(\Delta-\Delta_0)~\sn^2(\tau)]
\over\Delta+\Delta_0+(\Delta-\Delta_0)~\sn^2(\tau)}\right]~,
\label{phidotposksmall}
\earay
which is uniform up to corrections $\sim\Delta$. Note that in perfect
axisymmetry, $\Omvec_z$ and $d\phi/dt$ are independent of time, another
difference between the magnetic case, which is inherently triaxial,
and precession with an axisymmetric crust.
When the star is nearly axisymmetric,
\be
{d\phi\over dt}\simeq {L\over\Ibar}\left[1+\Delta
-(\Delta-\Delta_0)\sn^2(\tau)\right]~;
\ee
thus, $d\phi/dt$ oscillates at twice the precession frequency in
this limit. Associated with the time development of $d\phi/dt$
would also be variability of the pulsar spindown, at even harmonics
of the precession frequency. The amplitude of the main variation
would be of order $(\Delta-\Delta_0)P_0/P_p\sim\epsilon(P_0/P_p)^2$, 
where $\epsilon\sim(\Delta-\Delta_0)/2\Delta$ is the fractional
deviation from axisymmetry; successive harmonics would be smaller
by powers of $\epsilon$. For PSR 1828-11, we would have
$(\Delta-\Delta_0)P_0/P_p\sim 3\times 10^{-16}\epsilon$.
For comparison, electromagnetic spindown produces oscillations
with an amplitude $\sim \vert\Omhat_-\vert P_0/\tau_{sd}
\sim 1.3\times 10^{-13}\vert\Omhat_-\vert$, where $\tau_{sd}$ is
the spindown timescale, which is considerably
larger (Link \& Epstein 2001).

Applying Eqs. (\ref{eulerangles}) to $\bhat=\bhat_+\ehatplus
+\bhat_-\ehatminus$, and using Eqs. (\ref{alphapsipos}) we find
\baray
%\bhat_x&=&{\bhat_-\over\sqrt{\twofac}}
%\left[\sn(\tau)\cos\phi\sqrt{{2\Delta(1-\Delta_0)\over
%(\Delta+\Delta_0)(1-\Delta)}}
%-{\coseta(1+\Delta)~\dn(\tau)~\cn(\tau)\sin\phi\over\sqrt{\onefac}}
%\right]\nonumber\\& &
%+{\bhat_+\sineta(1-\Delta)\sin\phi\over\sqrt{\onefac}}
%\sqrt{\twofac}\nonumber\\
\bhat_y&=&{\bhat_-\over\sqrt{\twofac}}
\left[\sn(\tau)\sin\phi\sqrt{{2\Delta(1-\Delta_0)\over
(\Delta+\Delta_0)(1-\Delta)}}
+{\coseta(1+\Delta)~\dn(\tau)~\cn(\tau)\cos\phi\over\sqrt{\onefac}}
\right]\nonumber\\& &
-{\bhat_+\sineta(1-\Delta)\cos\phi\over\sqrt{\onefac}}
\sqrt{\twofac}\nonumber\\
%\bhat_z&=&\bhat_-\sineta(1-\Delta)~\cn(\tau)
%+{\bhat_+\coseta(1+\Delta)~\dn(\tau)\over\sqrt{\onefac}}~.
\label{bhatinertialpos}
\earay
No approximations have been made in Eq. (\ref{bhatinertialpos}); in
fact, there is also no explicit reference to magnetic distortions
here, and so these results apply to triaxial stars in general.
We note here that 
\baray
\bhat_+&=&{1\over\sqrt{2}}\left[1+{(C\cos 2\theta-I_B)\over
\sqrt{C^2-2I_BC\cos 2\theta+I_B^2}}\right]^{1/2}\nonumber\\
\bhat_-&=&-{1\over\sqrt{2}}\left[1-{(C\cos 2\theta-I_B)\over 
\sqrt{C^2-2I_BC\cos 2\theta+I_B^2}}\right]^{1/2}~,
\label{bhatpm}
\earay
and in the minimum energy state we can take $\theta\simeq\chi$
to lowest order in distortions. When $C$ is large compared with
$I_B$, these reduce to $\bhat_+\simeq\cos\chi$ and $\bhat\simeq
-\sin\chi$.

Pulse arrival times are found from the condition that
$\bhat_y=0$ if we assume that the observer is in the $x-z$ plane.
Let us simplify the notation by introducing the parameters
\baray
e^2&\equiv&{(\Delta-\Delta_0)(1+\Delta)\over (\Delta+\Delta_0)(1-\Delta)}
\nonumber\\
q_\pm&=&{1\pm\Delta\over\sqrt{\onefac}}~;
\earay
$e^2$ measures the departure from axisymmetry, and $q_\pm$ are one
to lowest order in the distortions. We can simplify further by
defining $\Sighat_\pm=q_\pm\Omhat_\pm$; then it follows that
$\cosigma^2+\sinsigma^2=1$.
In terms of these, the pulse arrival times are the solution to
\be
0={\bhat_-[\sqrt{1+e^2}\sn(\tau)\sin\phi+\cosigma
~\dn(\tau)\cn(\tau)\cos\phi]\over\sqrt{1+e^2\sn^2(\tau)}}
-\bhat_+\sinsigma\cos\phi\sqrt{1+e^2\sn^2(\tau)}~;
\ee
we can rewrite this as
\baray
\cos(\phi-\tau)&=&[\sin\tau-\sqrt{1+e^2}\sn(\tau)]\sin\phi
\nonumber\\& &
+\left\{\cos\tau-\cosigma\cn(\tau)\dn(\tau)
+{\bhat_+\over\bhat_-}\sinsigma[1+e^2\sn^2(\tau)]
\right\}\cos\phi~.
\earay
Define the functions
\baray
e^2F_S(\tau)&=&\sin\tau-\sqrt{1+e^2}\sn(\tau)\nonumber\\
e^2F_C(\tau)&=&\cos\tau-\cn(\tau)\dn(\tau)~;
\earay
then the pulse arrival times are the solutions of the
equation
\baray
\cos(\phi-\tau)=e^2F_S(\tau)\sin\phi
+\biggl\{e^2F_C(\tau)\cosigma+(1-\cosigma)\cos\tau
\nonumber\\
+{\bhat_+\over\bhat_-}\sinsigma[1+e^2\sn^2(\tau)]
\biggr\}\cos\phi~.
\label{arrivaltimespos}
\earay
In general, we can solve Eq. (\ref{arrivaltimespos}) numerically,
but a good analytic approximation can be found if the RHS
is small.

The terms on the RHS of Eq. (\ref{arrivaltimespos}) will
be small if the deviation from axisymmetry is small
and the rotation is nearly aligned with one of the
principal axes of the inertia tensor.
If we assume that RHS is small, then we can
develop an approximate solution by first substituting
$\phi=\phi-\tau+\tau$. If we let $\eta\equiv\phi-\tau$,
then we find the equation
\be
\cos\eta=\sin\eta\left[e^2F_S(\tau)\cos\tau-F(\tau)\sin\tau\right]
+\cos\eta\left[e^2F_S(\tau)\sin\tau+F(\tau)\cos\tau\right]~,
\label{arrivalexpandpos}
\ee
where
\be
F(\tau)\equiv e^2F_C(\tau)\cosigma+(1-\cosigma)\cos\tau
+{\bhat_+\over\bhat_-}\sinsigma[1+e^2\sn^2(\tau)]~.
\ee
We see that $F(\tau)$ is small only if both $e^2\ll 1$
and $1-\cosigma\simeq 1-\Omhat_+\ll 1$, so the expansion implied
here involves a pair of small parameters.
To zeroth order the solution to Eq. (\ref{arrivalexpandpos})
is $\eta=(2n+1/2)\pi$; writing the solution in general as
$\eta=(2n+1/2)\pi+\delta$ we find
\be
\tan\delta=-{[e^2F_S(\tau)\cos\tau-F(\tau)\sin\tau]\over
1-[e^2F_S(\tau)\sin\tau+F(\tau)\cos\tau]}~,
\ee
or, for small values of $\delta$,
\be
\delta\simeq [F(\tau)\sin\tau-e^2F_S(\tau)\cos\tau)]
\left[1+e^2F_S(\tau)\sin\tau+F(\tau)\cos\tau
+\cdots\right]~.
\label{approxsolnposksmall}
\ee
It should be apparent from Eq. (\ref{approxsolnposksmall}) that
arrival times should exhibit oscillations of decreasing
magnitude with frequencies $r\omega_p$, where $r$ is an integer.
Rigorously, we should expand the functions $F(\tau)$
and $F_S(\tau)$ in powers of $e^2$ as well, and
the hiden functions $\dn(\tau)$, $\cn(\tau)$ and
$\sn(\tau)$ should be expanded in $k^2$ (which is small
if $1-\Omhat_+^2$ is small: see Eq. [\ref{ktaudefpos}]).

For the triaxial case we have to consider, in addition to
$\delta$, the mapping between $t$ and $\phi$, which is
complicated by the fact that $d\phi/dt$ is time dependent.
Eq. (\ref{phidotposksmall}) implies that, to lowest 
nontrivial order in $e^2$,
\be
{\Ibar \over L}{d\phi\over dt}
\simeq 1+\Delta(1-e^2)+e^2\Delta\cos 2\tau~,
\label{phidotexpandposksmall}
\ee
which integrates to
\be
\phi\simeq \left[1+\Delta(1-e^2)\right]\int_0^t{dt~L\over\Ibar}
+{e^2\sin 2\tau\over 4\vert\coseta\vert}~.
\ee
%Since $\eta=\phi-\tau=\phi-\omega_pt\simeq
%\phi-2\Delta\vert\coseta\vert Lt/\Ibar$, we find that
and therefore
\be
\eta\simeq \left[1-\Delta(2\vert\coseta\vert-1+e^2)\right]
\int_0^t{dt~\over\Ibar}+{e^2\sin 2\tau\over 4\vert\coseta\vert}~.
\label{phiexpandposksmall} 
\ee 
The integral includes the effects of sinusoidal variations in
pulsar spindown, which may be evaluated using the formulae
in \S~\ref{spindown}.
Thus, the final solution for arrival times is
\be
\left[1-\Delta(2\vert\coseta\vert-1+e^2)\right]
\int_0^t{dt~L\over\Ibar}\simeq \left(2n+{1\over 2}\right)\pi
+\delta-{e^2\sin 2\tau\over 4\vert\coseta\vert}~.
\label{pulsarrivposksmall}
\ee
We see that the variation of $d\phi/dt$ with time
may introduce variation in pulse arrival times at
twice the precession frequency, a possibility that
only arises when the moment of inertia tensor is
nonaxisymmetric. Note that in the expansion in 
Eq. (\ref{phidotexpandposksmall}) we only kept terms
up to $\sim e^2$. Higher order terms would introduce
further variability (presumably at $4\omega_p$, $6\omega_p$
and so on).

%We shall present results of comparing a more systematic
%treatment of approximate solutions for the arrival times
%with data on PSR 1828-11 elsewhere (Akgun, Epstein
%\& Wasserman 2002).

\subsubsection{$I_B>0$ and $k^2>1$}
\label{app:pulsibposkbig}

For $k^2>1$, we choose $\Vvec_{\onehat}=\Vvec_+$,
$\Vvec_{\twohat}=\Vvec_2$ and $\Vvec_{\threehat}=
\Vvec_-$. We then find that
\baray
\cos\alpha&=&(1-\Delta)\dn(\tautil)\sqrt{{1-\Omhat_+^2\over
\onefac}}\nonumber\\
\tan\psi&=&{\rm sign}(\Omhat_-)~
{\cn(\tautil)\over\sn(\tautil)}\sqrt{{(1+\Delta)
(\Delta-\Delta_0)\over 2\Delta(1-\Delta_0)}}\nonumber\\
{d\phi\over dt}&=&{L\over\Ibar}\left[{\Delta-\Delta_0
+(\Delta+\Delta_0)\sn^2(\tautil)\over
(1+\Delta)(\Delta-\Delta_0)+(1-\Delta)(\Delta+\Delta_0)
\sn^2(\tautil)}\right]~.
\label{ibnegkbigparams}
\earay
For this case, we define $\Sighat_\pm$ as before, and
\be
\etil^2\equiv{(\Delta+\Delta_0)(1-\Delta)\over
(\Delta-\Delta_0)(1+\Delta)}~;
\ee
in terms of these variables, we find that
\be
\bhat_y={\bhat_+[{\rm sign}(\Omhat_-)\sqrt{1+\etil^2}\sn(\tautil)\sin\phi
+\sinsigma\dn(\tautil)\cos\phi]\over\sqrt{1+\etil^2\sn^2(\tautil)}}
-\bhat_-\cosigma\cos\phi\sqrt{1+e^2\sn^2(\tautil)}~,
\label{bhatykbig}
\ee
where $\bhat_\pm$ are given by Eq. (\ref{bhatpm}). Pulse arrival
times are found by solving $\bhat_y=0$. The equations are exactly
the same as for $k^2<1$ except for the replacements $e^2\to
\etil^2=1/e^2$, $\bhat_-\to{\rm sign}(\Omhat_-)\bhat_+$,
$\bhat_+\to\bhat_-$. $\cosigma\to\vert\sinsigma\vert$ and
$\sinsigma\to\cosigma$. Thus, if we define
$\eta=\phi-\tautil$, and also
\baray
\etil^2\Ftil_S(\tautil)&=&\sin\tautil-\sqrt{1+\etil^2}\sn(\tautil)
\nonumber\\
\etil^2\Ftil_C(\tautil)&=&\cos\tautil-\cn(\tautil)\dn(\tautil)
\nonumber\\
\Ftil(\tautil)&=&\etil^2\Ftil_C(\tautil)\vert\sinsigma\vert
+(1-\vert\sinsigma\vert)\cos\tau
+{\rm sign}(\Omhat_-){\bhat_-\over\bhat_+}\cosigma[1+\etil^2\sn^2(\tautil)]~,
\earay
then we find
\be
\cos\eta=\sin\eta\left[\etil^2\Ftil_S(\tautil)\cos\tautil-\Ftil(\tautil)
\sin\tautil\right]+\cos\eta\left[\etil^2\Ftil_S(\tautil)\sin\tautil
+\Ftil(\tautil)\cos\tautil\right]~.
\ee
As before, the approximate solution is $\eta=(2n+1/2)\pi+\delta$, 
where
\be
\tan\delta=-{[\etil^2\Ftil_S(\tautil)\cos\tautil-\Ftil(\tautil)
\sin\tautil]\over 1-[\etil^2\Ftil_S(\tautil)
+\Ftil(\tautil)\cos\tautil]}~,
\ee
which becomes
\be
\delta\simeq[\Ftil(\tautil)\sin\tautil-\etil^2\Ftil_S(\tautil)\cos\tautil]
[1+\etil^2\Ftil_S(\tautil)\sin\tautil+\Ftil(\tautil)\cos\tautil
+\cdots]
\label{approxsolposkbig}
\ee
for small $\delta$. This approximate solution is valid as long as $\vert\Omhat_-\vert
\simeq 1$ and $\etil^2\ll 1$. The restriction to small $\vert\Omhat_-\vert$
may seem a bit strange, until we recall that for $I_B>0$ and $\ktil^2=1/k^2<1$,
the spinning star is somewhat prolate rather than oblate. Thus, simple
solutions are expected for small amplitude around the axis of smallest moment
of inertia in this case.

The mapping from pulse phase to pulse arrival times requires the connection
between $\eta=\phi-\tautil$ and $t$. From the last of Eqs. (\ref{ibnegkbigparams})
we find that
\be
{\Ibar\over L}{d\phi\over dt}\simeq 1-\Delta(1-e^2)-e^2\Delta\cos 2\tautil~,
\ee
which integrates to 
\be
\phi\simeq \left[1-\Delta(1-e^2)\right]\int{dt~L\over\Ibar}
-{e^2\sin 2\tau\over 4\sqrt{1-\Omhat_+^2}}~,
\label{phimaposkbig}
\ee
when we use $\omega_p\simeq 2\Delta L\sqrt{1-\Omhat_+^2}/\Ibar$. Thus
we find that
\be
\eta=\phi-\tautil\simeq\left[1-\Delta\left(1+2\sqrt{1-\Omhat_+^2}-e^2\right)
\right]\int_0^t{dt~L\over\Ibar}-{e^2\sin 2\tautil\over 4\sqrt{1-\Omhat_+^2}}~,
\ee
so that pulses arrive at times
\be
\left[1-\Delta\left(1+2\sqrt{1-\Omhat_+^2}-e^2\right)
\right]\int_0^t{dt~L\over\Ibar}\simeq \left(2n+{1\over 2}\right)\pi
+\delta+{e^2\sin 2\tautil\over 4\sqrt{1-\Omhat_+^2}}~.
\label{pulsarrivposkbig}
\ee
As before, we see that to lowest order in $e^2$, the mapping from phase
to time introduces variation at $2\omega_p$. Retaining higher order terms
would introduce further variability at $4\omega_p$, etc.

\subsection{Pulse Arrival Times for $I_B<0$}
\label{app:pulsibneg}

For $I_B<0$, $\Vvec_{\onehat}=\Vvec_2$ and $\Vvec_{\twohat}
=\Vvec_-$, and Eq. (\ref{alphapsipos}) is replaced by
\baray
\cos\alpha&=&\dn(\tau)\sqrt{{[\Delta_0+\Delta^2
+\Delta(1+\Delta_0)(2\Omhat_+^2-1)](1+\Delta)
\over(\Delta+\Delta_0)[\onefac]}}\nonumber\\
\tan\psi&=&{\cn(\tau)\over\sn(\tau)}
\sqrt{{2\Delta(1-\Delta_0)\over(1-\Delta)
(\Delta+\Delta_0)}}~,
\label{alphapsineg}
\earay
and Eq. (\ref{phidotposksmall}) is replaced by
\be
{d\phi\over dt}={L\over\Ibar}\left[{2\Delta+(\Delta_0-\Delta)\sn^2(\tau)
\over 2\Delta(1-\Delta_0)+(\Delta_0-\Delta)(1+\Delta)\sn^2(\tau)}\right]
~.
\label{phidotneg}
\ee
In this case, we define
\baray
\cosigma&=&\sqrt{(1+\Delta)
[\Delta_0+\Delta^2+\Delta(1+\Delta_0)(2\Omhat_+^2-1)]\over
(\Delta+\Delta_0)[\onefac]}\nonumber\\
\sinsigma&=&\Omhat_-\sqrt{2\Delta(1-\Delta_0)(1-\Delta)\over
(\Delta+\Delta_0)[\onefac]}\nonumber\\
e^2&=&{(\Delta_0-\Delta)(1+\Delta)\over 2\Delta(1-\Delta_0)}~;
\earay
then
\be
\bhat_y=-{\bhat_-[\cn(\tau)\sin\phi-\cosigma\sqrt{1+e^2}\dn(\tau)\sn(\tau)
\cos\phi]\over\sqrt{1+e^2\sn^2(\tau)}}
-\bhat_+\sinsigma\cos\phi\sqrt{1+e^2\sn^2(\tau)}~,
\ee
where $\bhat_\pm$ are given by Eq. (\ref{bhatpm}). Pulse arrival
times are determined from the condition $\bhat_y=0$:
\baray
\sin(\phi-\tau)&=&[\cos\tau-\cn(\tau)]\sin\phi
\nonumber\\& &
+\left\{\cosigma\sqrt{1+e^2}\sn(\tau)\dn(\tau)-\sin\tau
+{\bhat_+\over\bhat_-}\sinsigma[1+e^2\sn^2(\tau)]\right\}
\cos\phi~.\nonumber\\
\earay
Define the functions
\baray
e^2G_C(\tau)&=&\cos\tau-\cn(\tau)\nonumber\\
e^2G_S(\tau)&=&\sqrt{1+e^2}\sn(\tau)\dn(\tau)-\sin\tau\nonumber\\
G(\tau)&=&\cosigma e^2G_S(\tau)-(1-\cosigma)\sin\tau
+{\bhat_+\over\bhat_-}\sinsigma[1+e^2\sn^2(\tau)]~;
\earay
then the pulse arrival times are solutions of the equation
\be
\sin\eta=[e^2G_C(\tau)\cos\tau-G(\tau)\sin\tau]\sin\eta
+[e^2G_C(\tau)\sin\tau+G(\tau)\cos\tau]\cos\eta~,
\ee
where $\eta\equiv\phi-\tau$. The solution is $\eta=2\pi n
+\delta$, where 
\be
\tan\delta={e^2G_C(\tau)\sin\tau+G(\tau)\cos\tau
\over 1-[e^2G_C(\tau)\cos\tau-G(\tau)\sin\tau}~,
\ee
or, for small $\delta$,
\be
\delta\simeq [e^2G_C(\tau)\sin\tau+G(\tau)\cos\tau]
[1+e^2G_C(\tau)\cos\tau-G(\tau)\sin\tau]~.
\label{approxsolneg}
\ee
Once again, it is necesary that both $e^2\ll 1$ and
$\sqrt{1-\Omhat_+}\ll 1$ for the approximation to be
valid.

For this case, the mapping from phase to arrival times
that results from expanding Eq. (\ref{phidotneg}) leads
to exactly the same result as for $I_B>0$ and $k^2<1$, namely
\be
\eta\simeq \left[1-\Delta\left(2\coseta-1+e^2\right)\right]
\int_0^t{dt~L\over\Ibar}+{e^2\sin 2\tau\over 4\vert\Omhat_+\vert}~.
\ee
In this case, then, pulses arrive when
\be
\left[1-\Delta\left(2\coseta-1+e^2\right)\right] 
\int_0^t{dt~L\over\Ibar}\simeq 2n\pi+\delta-{e^2\sin 2\tau\over 4\vert\Omhat_+\vert}~.  
\label{pulsarrivneg}
\ee

\end{document}